\documentclass[aps,pra,onecolumn,preprintnumbers,amsmath,amssymb,superscriptaddress,nofootinbib]{revtex4-2}

\usepackage[utf8]{inputenc}
\usepackage[T1]{fontenc}
\usepackage[english]{babel}
\usepackage{graphicx}
\usepackage{xcolor}
\usepackage[hidelinks]{hyperref}
\usepackage{listings}
\usepackage{quantikz}
\usepackage{mathtools}
\usepackage{booktabs}
\usepackage{mdframed}

\definecolor{codegreen}{rgb}{0,0.6,0}
\definecolor{codegray}{rgb}{0.5,0.5,0.5}
\definecolor{codepurple}{rgb}{0.58,0,0.82}
\definecolor{backcolour}{rgb}{0.95,0.95,0.92}
\definecolor{darkblue}{rgb}{0.0, 0.0, 0.55}

\hypersetup{
    colorlinks=true,       
    linkcolor=darkblue,    
    citecolor=codegreen,    
    urlcolor=codepurple,     
    bookmarksnumbered=true, 
    pdfstartview={FitH}, 
    pdftitle={TensorCircuit-NG Whitepaper}, %
    pdfauthor={Shi-Xin Zhang et al.},       %
}
\lstdefinestyle{mystyle}{
    backgroundcolor=\color{backcolour},   
    commentstyle=\color{codegreen},
    keywordstyle=\color{magenta},
    numberstyle=\tiny\color{codegray},
    stringstyle=\color{codepurple},
    basicstyle=\ttfamily\footnotesize,
    breakatwhitespace=false,         
    breaklines=true,                 
    captionpos=b,                    
    keepspaces=true,                 
    numbers=none,                    
    numbersep=5pt,                  
    showspaces=false,                
    showstringspaces=false,
    showtabs=false,                  
    tabsize=2,
    frame=none
}
\usepackage{xcolor}
\usepackage[normalem]{ulem}
\usepackage{realboxes}
\definecolor{bkgd}{RGB}{240,242,246}
\definecolor{orange-red}{rgb}{1.0, 0.27, 0.0}

\newcommand{\tc}{TensorCircuit-NG} 

\newcommand{\fancylink}[2]{\colorbox{bkgd}{\color{orange-red}\href{#1}{\sf {#2}}}}
\newcommand{\rooturl}{https://github.com/tensorcircuit/tensorcircuit-ng/blob/master/}
\newcommand{\docurl}{https://tensorcircuit-ng.readthedocs.io/en/latest/}

\begin{document}

\title{TensorCircuit-NG: A Universal, Composable, and Scalable Platform for Quantum Computing and Quantum Simulation}

\author{Shi-Xin Zhang}
\email{shixinzhang@iphy.ac.cn}
\affiliation{Institute of Physics, Chinese Academy of Sciences, Beijing 100190, China}
\author{Yu-Qin Chen}
\affiliation{Graduate School of China Academy of Engineering Physics, Beijing 100193, China}
\author{Weitang Li}
\affiliation{School of Science and Engineering, The Chinese University of Hong Kong, Shenzhen, Guangdong 518172, China}
\author{Jiace Sun}
\affiliation{Division of Chemistry and Chemical Engineering, California Institute of Technology, Pasadena, CA 91125, USA}
\author{Wei-Guo Ma}
\affiliation{Institute of Physics, Chinese Academy of Sciences, Beijing 100190, China}
\affiliation{University of Chinese Academy of Sciences, Beijing 100049, China}
\author{Pei-Lin Zheng}
\affiliation{Future Research Lab, China Mobile Research Institute, Beijing 100053, China}
\author{Yu-Xiang Huang}
\affiliation{Institute of Physics, Chinese Academy of Sciences, Beijing 100190, China}
\affiliation{University of Chinese Academy of Sciences, Beijing 100049, China}
\author{Qi-Xiang Wang}
\affiliation{University of Chinese Academy of Sciences, Beijing 100049, China}
\author{Hui Yu}
\affiliation{Institute of Physics, Chinese Academy of Sciences, Beijing 100190, China}
\author{Zhuo Li}
\affiliation{School of Computer Science and Artificial Intelligence, Zhengzhou University, Zhengzhou 450001, China}
\author{Xuyang Huang}
\affiliation{Institute for Quantum Science and Technology, Shanghai University, Shanghai 200444, China}
\author{Zong-Liang Li}
\affiliation{Institute of Physics, Chinese Academy of Sciences, Beijing 100190, China}
\affiliation{University of Chinese Academy of Sciences, Beijing 100049, China}
\author{Zhou-Quan Wan}
\affiliation{Center for Computational Quantum Physics, Flatiron Institute, New York, NY 10010, USA}
\author{Shuo Liu}
\affiliation{Department of Physics, Princeton University, Princeton, NJ 08544, USA}
\author{Jiezhong Qiu}
\affiliation{Hangzhou Institute of Medicine, Chinese Academy of Sciences, Hangzhou 310018, China}
\author{Jiaqi Miao}
\affiliation{Department of Physics and Astronomy, Rice University, Houston, TX 77005, USA}
\affiliation{Applied Physics Graduate Program, Smalley-Curl Institute, Rice University, Houston, TX 77005, USA}
\author{Zixuan Song}
\affiliation{Department of Computer Science and Engineering, Washington University in St. Louis, St. Louis, MO 63130, USA}
\author{Yuxuan Yan}
\affiliation{Center for Quantum Information, Institute for Interdisciplinary
Information Sciences, Tsinghua University, Beijing 100084, China}
\author{Kazuki Tsuoka}
\affiliation{Department of Chemistry, University of Tokyo, 7-3-1 Hongo, Bunkyo-ku, Tokyo 113-0033, Japan}

\author{Pan Zhang}
\affiliation{Institute of Theoretical Physics, Chinese Academy of Sciences, Beijing 100190, China}
\affiliation{School of Fundamental Physics and Mathematical Sciences,
Hangzhou Institute for Advanced Study, UCAS, Hangzhou 310024, China}
\author{Lei Wang}
\affiliation{Institute of Physics, Chinese Academy of Sciences, Beijing 100190, China}
\author{Heng Fan}
\affiliation{Institute of Physics, Chinese Academy of Sciences, Beijing 100190, China}
\affiliation{University of Chinese Academy of Sciences, Beijing 100049, China}
\affiliation{Beijing Academy of Quantum Information Sciences, Beijing 100193, China}
\author{Chang-Yu Hsieh}
\affiliation{College of Pharmaceutical Sciences, Zhejiang University, Hangzhou, 310058, China}
\author{Hong Yao}
\affiliation{Institute for Advanced Study, Tsinghua University, Beijing 100084, China}
\author{Tao Xiang}
\affiliation{Institute of Physics, Chinese Academy of Sciences, Beijing 100190, China}
\affiliation{University of Chinese Academy of Sciences, Beijing 100049, China}
\date{\today}

\begin{abstract}
We present \tc, a next-generation quantum software platform designed to bridge the gap between quantum physics, artificial intelligence, and high-performance computing. Moving beyond the scope of traditional circuit simulators, \tc\ establishes a unified, tensor-native programming paradigm where quantum circuits, tensor networks, and neural networks fuse into a single, end-to-end differentiable computational graph. Built upon industry-standard machine learning backends (JAX, TensorFlow, PyTorch), the framework introduces comprehensive capabilities for approximate circuit simulation, analog dynamics, fermion Gaussian states, qudit systems, and scalable noise modeling. To tackle the exponential complexity of deep quantum circuits, \tc\ implements advanced distributed computing strategies, including automated data parallelism and model-parallel tensor network slicing. We validate these capabilities on GPU clusters, demonstrating a near-linear speedup in distributed variational quantum algorithms. \tc\ enables flagship applications, including end-to-end QML for CIFAR-100 computer vision, efficient pipelines from quantum states to neural networks via classical shadows, and differentiable optimization of tensor network states for many-body physics.

\end{abstract}

\maketitle
\tableofcontents

\section{Introduction}
The domain of computational science is undergoing a fundamental paradigm shift: from discrete, static simulation to differentiable, data-driven discovery. This transition demands a new generation of computational infrastructure where physical models are not merely executed, but are intimately integrated with gradient-based optimization and machine learning (ML) pipelines, a philosophy exemplified by the original TensorCircuit framework~\cite{Zhang2022tc}.

While various quantum software ecosystems have successfully served the community for algorithm verification and educational purposes~\cite{Javadi-Abhari2024, cirq_developers_2025_16867504, Johansson2012, ProjectQ2018, TFQ2020, Yao2020, Qulacs2021, Qibo2021}, the computational landscape is rapidly evolving. The field is shifting from standalone circuit simulators to integrated platforms capable of deeply bridging quantum physics modeling, artificial intelligence (AI), and high-performance computing (HPC). This transition is driven by the necessity to treat quantum circuits not merely as isolated instruction sequences, but as differentiable components within complex, data-driven scientific workflows. \tc\ addresses this demand by establishing a unified framework where digital logic, analog dynamics, tensor networks, and neural networks are fused under a backend-agnostic programming paradigm.
This architecture enables a seamless transition from simulating proof-of-concept protocols to discovering hybrid algorithms and modeling many-body physics at scale.

\subsection{The Evolution of TensorCircuit}
Since its inception, TensorCircuit has established itself as a pioneering framework by
bridging the gap between quantum simulation and modern ML
infrastructure~\cite{Zhang2022tc}.
The original version of TensorCircuit focused on constructing a high-performance tensor network engine~\cite{Markov2008, Pan2022} built
on top of industry-standard backends (TensorFlow~\cite{10.5555/3026877.3026899}, JAX~\cite{Frosting2018}, PyTorch~\cite{10.5555/3454287.3455008}).
By fundamentally integrating ML engineering paradigms—automatic
differentiation (AD)~\cite{Rumelhart1986, Baydin2018, Liao2019, Zhang2019b, Wan2020}, just-in-time (JIT) compilation, and vectorized parallelism (VMAP)—it transformed quantum circuit simulation from rigid
state-vector updates into efficient, differentiable computational graphs.
This framework has been widely adopted by the community, empowering a global user base spanning over 50 research institutions and supporting more than 150
academic publications.
It has become a standard tool for cutting-edge research in variational quantum algorithms (VQA)~\cite{Cerezo2020breview,Bharti2021review} and quantum machine learning (QML)~\cite{Biamonte2017nature}, proven to
accelerate discovery in the noisy intermediate-scale quantum era \cite{Preskill2018}.

However, the frontier of quantum research has since significantly expanded \cite{Preskill2025}.
In this Next Generation release, we transition \tc\ from a circuit simulator to a
comprehensive quantum science and technology platform.
This evolution is driven by the necessity to support a diverse spectrum of
scientific inquiries: from condensed matter physicists modeling continuous-time
many-body dynamics and control engineers optimizing analog pulses, to ML practitioners embedding quantum layers into complex, distributed classical pipelines.
\tc\ retains the robust differentiable engine of its predecessor while
introducing a modular ecosystem designed for quantum computing and quantum physics with scalable distributed computing capability.

\subsection{Foundations Recap}
\begin{mdframed}
\textbf{Reference Script: }
\fancylink{\rooturl examples/ng_whitepaper/IB_batch_vqe_tfim.py}{\sf IB\_batch\_vqe\_tfim.py}
\end{mdframed}

The core architecture of TensorCircuit, as detailed in our original work~\cite{Zhang2022tc}, relies on a backend-agnostic tensor network engine. By decoupling the physics logic from the computational backend, TensorCircuit leverages the JIT, VMAP, and AD capabilities of the underlying backend.
In \tc, these foundational features remain the bedrock. The efficient tensor contraction engine provides the necessary performance and gradient information that power the advanced physics modeling and distributed computing features introduced in this work.

The core philosophy of TensorCircuit established in its first release is the fusion of quantum physics primitives with machine learning paradigms. To illustrate this foundation, we revisit the variational quantum eigensolver (VQE)~\cite{Peruzzo2014, Tilly2021} for the one-dimensional transverse field Ising model (TFIM).

VQE represents a prototypical hybrid quantum-classical algorithm designed to find the ground state energy of a Hamiltonian $H$.
It operates by parameterizing a quantum state $|\psi(\theta)\rangle$ via an ansatz circuit and minimizing the expectation value $E(\theta) = \langle \psi(\theta)|H|\psi(\theta)\rangle$ using a classical optimizer.
For the 1D TFIM, the Hamiltonian is defined as $H = - \sum_{\langle i,j \rangle} Z_i Z_j - g \sum_i X_i$.
In \tc, this simulation pipeline is constructed in three logical stages: definition (circuit primitives), transformation (AD, JIT, VMAP), and execution (optimization loop).

\textbf{Environment Setup}.
The simulation workflow begins with the global environment configuration.
This design allows users to tailor the underlying engine to their specific hardware and accuracy requirements without changing the physics code.
For this VQE task, we configure the framework to use JAX backend, the \texttt{cotengra} library for optimized tensor contraction paths~\cite{Gray2021}, and double-precision complex numbers:

\begin{lstlisting}[language=Python]
import tensorcircuit as tc

# 1. Backend: Use JAX for JIT and VMAP support
K = tc.set_backend("jax")

# 2. Contractor: Use 'cotengra' with default configuration for optimized tensor network contraction paths
tc.set_contractor("cotengra")

# 3. Precision: Use double precision for accurate scientific simulation
tc.set_dtype("complex128")
\end{lstlisting}

\textbf{Primitives: Tensor-Native Circuit Definition}.
Unlike traditional state-vector simulators that update a large global state array in place, \tc\ constructs a symbolic computational graph.
Quantum states, circuits and expectations are all treated as tensor networks, and quantum gates applications are modeled as local tensor contractions.
We define the parameterized ansatz and the energy expectation value purely using these backend-agnostic primitives:

\begin{lstlisting}[language=Python]
def tfim_energy(param, g=1.0, n=10):
    c = tc.Circuit(n)
    # --- Ansatz Construction ---
    for i in range(n):
        c.rx(i, theta=param[0, i]) # Layer 1: Single qubit rotations
    for i in range(n - 1):
        c.rzz(i, i+1, theta=param[1, i]) # Layer 2: Entangling ZZ gates
    
    # --- Expectation Calculation ---
    e = 0.0
    # Interaction term <Z_i Z_{i+1}>
    for i in range(n - 1):
        e -= c.expectation_ps(z=[i, i+1])
    # Transverse field term <X_i>
    for i in range(n):
        e -= g * c.expectation_ps(x=[i])
    return tc.backend.real(e)
\end{lstlisting}

\textbf{Paradigms: AD, JIT and VMAP}.
The static Python function above is transformed into a high-performance execution kernel using ML paradigms. 
\begin{itemize}
    \item \textbf{AD:} Instead of using parameter-shift rules which require $2k$ circuit evaluations for $k$ parameters, TensorCircuit utilizes reverse-mode AD (backpropagation) through the tensor network. This allows gradients to be computed with a computational cost independent of the number of parameters.
    \item \textbf{JIT:} The tensor contraction path and the gradient graph are traced once and compiled into optimized machine code (XLA for JAX/TF) for the target hardware. This removes Python overhead and enables operator fusion.
    \item \textbf{VMAP:} \tc\ automatic vectorization can handle batch processing. This feature promotes the batch dimension (e.g., random seeds or batched inputs) directly into the tensor operations, allowing thousands of independent circuit instances to be executed simultaneously. This eliminates the overhead of Python loops and maximizes device utilization.
\end{itemize}

The transformation from a physics definition to an optimized training step is achieved concisely as:

\begin{lstlisting}[language=Python]
# 1. Automatic Differentiation (AD)
# Compute value and gradients via reverse-mode AD (Backpropagation)
vqe_grad = K.value_and_grad(tfim_energy)

# 2. Vectorized Parallelism (VMAP)
# Automatically batch the computation over the first argument (params)
# This enables parallel optimization of independent random seeds
vqe_batch = K.vmap(vqe_grad, vectorized_argnums=0)

# 3. Just-In-Time Compilation (JIT)
# Compile the entire batched gradient graph into optimized XLA machine code
vqe_step = K.jit(vqe_batch, static_argnums=(2,))

# --- Execution ---
# params shape: [batch_size, layers, n]
# Executed in parallel on GPU/TPU
energies, grads = vqe_step(params, 1.0, 10) 
\end{lstlisting}

This architecture ensures that \tc\ is not merely simulating a quantum circuit step-by-step, but is executing a holistic computational graph optimized for modern hardware. This ``tensor-in, tensor-out'' design allows various \tc\ features such as analog simulation and classical shadows to immediately inherit these acceleration capabilities.

\subsection{The Next-Generation Leap: Key Innovations}

This manuscript details the major advancements in \tc, structured around four pillars that collectively redefine the capability scope of quantum software:

\begin{enumerate}
    \item \textbf{Deep Interoperability and Ecosystem:}
    We introduce native interfaces and object-oriented layers that allow quantum circuits to function as standard, end-to-end differentiable modules within PyTorch and Keras pipelines.
    Furthermore, we implement robust translation utilities for tensor network objects, enabling bidirectional conversion between \tc\ and other ecosystem tools (e.g., \texttt{quimb}~\cite{Gray2018}, \texttt{TensorNetwork}~\cite{Roberts2019}, \texttt{TeNPy}~\cite{Hauschild2018, Hauschild2024}) to leverage specialized external tools.

    \item \textbf{Complex Physical Modeling:}
    Moving beyond abstract ideal qubits, we significantly expand the scope of physical representation to capture the complexity of real-world quantum systems.
    We introduce native support for high-dimensional \textbf{qudit} systems ($d \ge 3$) and \textbf{fermion} Gaussian states, integrated with a high-level \texttt{Lattice} API for constructing arbitrary geometric topologies.
    Furthermore, to model dynamic and open systems, we incorporate advanced solvers for continuous \textbf{time evolution} and a highly customizable noise configuration system, enabling the faithful simulation of many-body dynamics under realistic experimental conditions.

    \item \textbf{Comprehensive Simulation Paradigms:}
    We significantly expand the simulation backend to support diverse regimes:
    (i) Analog simulation for time-dependent Hamiltonians and pulse-level control;
    (ii) Stabilizer simulation for large-scale Clifford circuits and error correction codes;
    (iii) Approximate simulation leveraging matrix product states (MPS) for limited-entangled dynamics.

    \item \textbf{HPC-Ready Scalability:}
    Addressing the memory and runtime bottlenecks of wide and deep circuits, we implement advanced distributed computing strategies.
    These include data-parallel evaluation of observables for variational algorithms and model-parallel tensor slicing techniques that distribute large-scale contractions across multi-GPU clusters.
\end{enumerate}

\subsection{Availability and Community}
 \tc\ is open-sourced under the Apache 2.0 license. The software is available on the Python Package Index and can be installed using the command {\sf pip install tensorcircuit-ng}.
The development of \tc\ is open-sourced and centered on GitHub: \fancylink{https://github.com/tensorcircuit/tensorcircuit-ng}{\tc\ Repository}. 

\tc\ is the actively maintained official version and a fully compatible successor to TensorCircuit.
We emphasize that this transition is designed to be frictionless: \tc\ serves as a strict drop-in replacement.
Users can migrate to this next-generation engine without modifying a single line of their existing code.
By preserving the identical API structure and import namespace, we ensure that the existing research scripts and community tutorials remain fully operational, allowing users to effortlessly adopt the new infrastructure and immediately benefit from the enhanced performance and physical modeling capabilities.

We believe that documentation is as important as the code itself. \tc\ features comprehensive online \fancylink{\docurl}{documentation}, including detailed API references and theoretical backgrounds. Furthermore, the repository hosts a rich collection of over 130 executable examples (some referenced throughout this paper), serving as a practical cookbook for users to quickly prototype ideas.

\tc\ is envisioned as a community-driven platform. We actively encourage feedback and contributions from the quantum science community. Whether through reporting issues, engaging in GitHub Discussions, or submitting Pull Requests for new features, researchers are invited to co-create the future of \tc\ with us.

\subsection{AI-Native Development}
\tc\ is engineered not only for human developers but also for the era of AI-assisted programming. To bridge the gap between static documentation and practical coding, we introduce the \fancylink{https://deepwiki.com/tensorcircuit/tensorcircuit-ng}{TensorCircuit-NG DeepWiki}. This AI-powered assistant allows users to explore the library's functionality through natural language queries, providing instant guidance. We also recognize that modern research workflows increasingly rely on AI coding agents (e.g., Cursor, Claude Code, Codex, Antigravity) to accelerate algorithm implementation. To facilitate this, the repository is structured to serve as a semantic context for large language models. We provide a dedicated \texttt{AGENTS.md}, which functions as a specialized handbook for AI agents, defining coding standards, API patterns, and best practices to ensure that the generated code is idiomatic and performant. Furthermore, the repository includes \texttt{llm\_experience.md}, which encapsulates specific physics heuristics and algorithmic protocols, effectively a knowledge base of distilled experiences curated for AI consumption.

To maximize the efficacy of these tools, we strongly recommend an in-repository development workflow, even for users focusing solely on application-level scripting. Instead of writing scripts in an empty directory, users are advised to clone the repository and establish a local branch. Opening this environment in an AI-integrated IDE grants the agent immediate access to the rich context of over 100 executable scripts in \texttt{examples} and extensive test cases in \texttt{tests}. By anchoring the AI's generation in these ground-truth examples, this workflow significantly mitigates hallucination, enabling researchers to implement complex quantum algorithms entirely through natural language prompts.

\section{Architecture and Unified Paradigm}

\tc\ is architected to dismantle the barriers between quantum physics simulation, hardware execution, and machine learning. By reimagining the quantum software stack as a differentiable physics engine, we introduce the unified quantum programming paradigm.
This paradigm is not merely a user interface convenience; it is the direct consequence of our tensor-native architectural philosophy. As illustrated in Table~\ref{tab:unified_programming}, \tc\ unifies the diverse landscape of infrastructures and interfaces, allowing researchers to define the physics once and deploy it anywhere—from a local laptop CPU to GPU clusters.

\begin{table}[h]
\centering
\caption{TensorCircuit-NG: Unified Quantum Programming. This architecture enables seamless switching between diverse computational and physical resources and interfaces.}
\label{tab:unified_programming}
\begin{tabular*}{\linewidth}{p{0.47\linewidth} @{\extracolsep{\fill}} p{0.47\linewidth}}
\toprule
\textbf{Unified Backends} & \textbf{Unified Devices} \\
Seamless switching between JAX, TensorFlow, PyTorch, NumPy, and CuPy. 
& Transparent execution on CPUs, NVIDIA GPUs, and Google TPUs without code modification. \\
\addlinespace[2ex] 

\textbf{Unified Providers} & \textbf{Unified Resources} \\
A single API to access QPUs from different vendors alongside local simulators. 
& Consistent workflow across local laptops, cloud instances, and HPC clusters. \\
\addlinespace[2ex]

\textbf{Unified Interfaces} & \textbf{Unified Engines} \\
Zero-code-change transition from high-precision simulation verification to experimental hardware deployment. 
& Switchable solvers: ideal state-vector, noisy density matrix, matrix product approximation, analog, or stabilizers. \\
\addlinespace[2ex]

\textbf{Unified Representations} & \textbf{Unified Objects} \\
Bidirectional translation between different frameworks for circuits and tensor networks. 
& First-class support for hybrid objects: neural networks, tensor networks, and quantum circuits interact natively. \\
\bottomrule
\end{tabular*}
\end{table}

\begin{figure}[t]
    \centering
\includegraphics[width=0.75\textwidth]{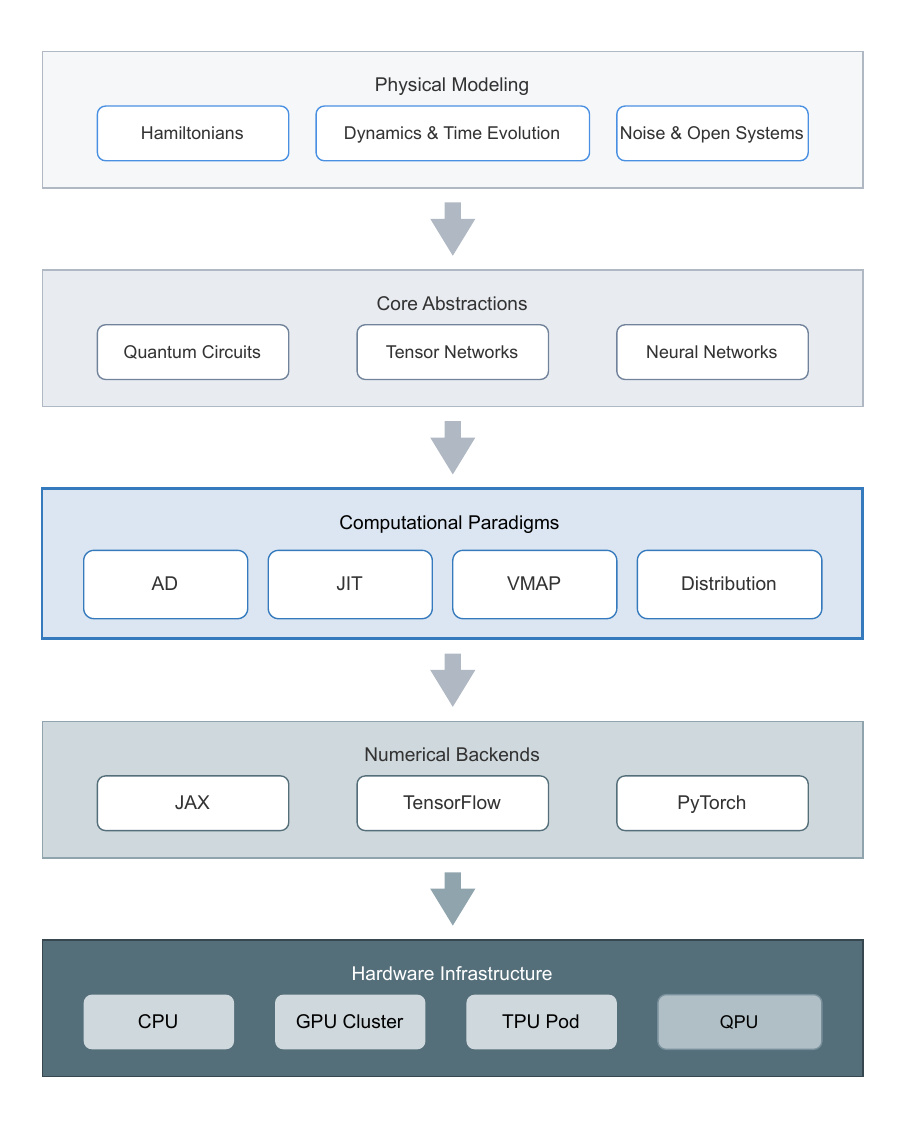}
    \caption{\textbf{The Hierarchical Software Architecture of \tc.} 
    The platform is structured into five distinct layers, bridging raw hardware acceleration with high-level physical modeling.
    \textbf{(Bottom)} The \textit{Hardware Infrastructure} layer abstracts diverse compute resources, seamlessly dispatching workloads to CPUs, GPU clusters, TPUs, or quantum processors (QPUs).
    \textbf{(Second)} The \textit{Numerical Backends} layer unifies industry-standard ML frameworks (JAX, TensorFlow, PyTorch) via a backend-agnostic dispatch interface.
    \textbf{(Third)} The \textit{Computational Paradigms} engine serves as the core transformation layer, injecting AD, JIT, VMAP, and distributed computing capabilities into all upstream objects.
    \textbf{(Fourth)} \textit{Core Abstractions} unify quantum circuits, tensor networks, and neural networks into composable, differentiable tensors.
    \textbf{(Top)} The \textit{Physical Modeling} layer provides domain-specific modules for constructing Hamiltonians, simulating open system noise, and evolving many-body dynamics.}
    \label{fig:architecture}
\end{figure}

The architectural design of \tc\ is visualized in Fig.~\ref{fig:architecture}. It adopts a strict hierarchical approach designed to decouple physical modeling from the underlying computational engine. At the foundation, the framework abstracts the complexity of heterogeneous hardware through a unified backend interface. The central innovation lies in the middle layer: a paradigm-driven engine that automatically transforms high-level physical definitions (Hamiltonians, Circuits) into optimized, differentiable, and distributed execution kernels. This layered architecture ensures that top-level applications, such as QML or many-body dynamics, automatically inherit the performance benefits of the underlying AI infrastructure without requiring domain-specific optimization.

\subsection{Infrastructure Unification: The Dual-Layer Design}

The capability to seamlessly switch software backends and hardware stems from our  \textbf{physics-compute decoupling} design.

\begin{itemize}
    \item \textbf{The Physics Layer (Frontend):} Defines abstractions such as \texttt{Circuit}, \texttt{Lattice}, and \texttt{Hamiltonian}. This layer is pure Python and backend-agnostic.
    \item \textbf{The Runtime Layer (Backend):} Handles numerical execution. Through a unified dispatch interface, the framework translates physical logic into backend-specific tensor operations.
\end{itemize}

With this design paradigm, users can prototype a variational algorithm using NumPy for debugging, and with a single line of configuration, migrate the exact same code to JAX for massive parallel training on accelerators.

\subsection{Representation Unification: Tensor-Native Philosophy}

The unification of physical objects and numerical data is achieved because everything is a Tensor in \tc. Unlike frameworks that encapsulate quantum states in opaque wrapper classes, \tc\ treats quantum states, gates, and noise channels as first-class tensors directly.

This \textbf{tensor-native} foundation unifies the simulation of quantum mechanics with the paradigm of differentiable programming. We adhere to the principle that any physical parameter defining the system should be differentiable. By utilizing automatic differentiation through the unified tensor representation, \tc\ allows gradients to flow not just through circuit parameters, but through the entire modeling pipeline. This includes geometric parameters, Hamiltonian coefficients, control parameters and noise strength.
This capability transforms quantum simulation from a tool for verification (calculating observables given a model) into a tool for inverse design (optimizing the model to achieve target properties).

\subsection{Ecosystem Unification: Hybrid and Interoperable}

\tc\ extends the unification philosophy beyond its internal infrastructure to the broader software landscape, dismantling the boundaries between quantum and classical ecosystems.

\begin{itemize}
    \item \textbf{Hybrid Fluidity:} 
    By complying with standard tensor interfaces, \tc\ unifies quantum circuits with classical neural networks and tensor networks. A quantum circuit is simply a differentiable layer within the computational graph, allowing it to be composed arbitrarily with modules from PyTorch, TensorFlow, or JAX. This enables the training of truly hybrid models where backpropagation optimizes classical and quantum weights simultaneously.
    
    \item \textbf{Radical Interoperability:} 
    Recognizing the diversity of the quantum community, \tc\ unifies disparate tools into a cohesive workflow. We provide low-overhead, bidirectional translation to external formats. Users can import ansatz libraries from Qiskit~\cite{Javadi-Abhari2024}, export circuits to OpenQASM~\cite{Cross2017}, or interface with tensor network libraries like TeNPy~\cite{Hauschild2018, Hauschild2024} and Quimb~\cite{Gray2018}. This allows researchers to leverage the best-in-class tools from the entire ecosystem without being locked into a single framework.
\end{itemize}

\subsection{Engine Unification: Multi-Paradigm Simulation Solvers}

Finally, the unified interface for simulation engines allows users to trade off precision for scale without rewriting code. The same abstract circuit object can be executed via:
\begin{itemize}
    \item \textbf{Exact Engine:} Full tensor network contraction for verification.
    \item \textbf{Noisy Engine:} Monte Carlo trajectories or density matrix evolution for open systems.
    \item \textbf{MPS Engine:} Approximate simulation with tradeoff between precision and scale.
    \item \textbf{Stabilizer Engine:} Large scale simulation with tableau formalism for a subset of quantum operations.
\end{itemize}

With this unified architecture established, the following sections will delve into specific new features and applications, starting with how \tc\ better models the physical world.

\section{Deep Framework Interoperability and Ecosystem}

\subsection{Native Backend Interfaces}

\begin{mdframed}
\textbf{Reference Script: }
\fancylink{\rooturl examples/ng_whitepaper/IIIA_interface_demo.py}{\sf IIIA\_interface\_demo.py}
\end{mdframed}

\tc\ is fundamentally architected as a machine-learning-native framework, built upon a backend-agnostic tensor engine. Users can globally switch the computation backend with a single command \texttt{tc.set\_backend}, or efficiently isolate execution contexts using the \texttt{tc.runtime\_backend} context manager. This flexibility ensures that users can always leverage the specific strengths—such as JAX's JIT compilation speed or PyTorch's dynamic graph capabilities and ecosystems—of the underlying framework.

However, advanced research often necessitates a \textit{mixed-backend} computational graph, where different components of a single differentiable pipeline utilize different frameworks. A prime example is executing a computationally intensive quantum circuit using the high-performance JAX backend (leveraging XLA compilation) while managing the data loading and optimization loop within the widely-used PyTorch ecosystem.

To enable this, \tc\ introduces native backend interfaces. These interfaces wrap a quantum function executed on an internal backend and expose it as a native, differentiable operator to a host framework.
Crucially, this bridge is built upon the DLPack standard, enabling zero-copy tensor conversion between frameworks. The interface automatically manages the memory sharing and connects the automatic differentiation graphs, ensuring that gradients propagate seamlessly across the framework boundary without user intervention.

\begin{lstlisting}[language=Python]
import torch
import tensorcircuit as tc

# 1. Define a backend-agnostic quantum function
def universal_circuit(params):
    c = tc.Circuit(2)
    c.h(0)
    c.rx(0, theta=params[0])
    c.ry(1, theta=params[1])
    c.cx(0, 1)
    return tc.backend.real(c.expectation([tc.gates.x(), [0]]))

# 2. Cross-Framework Wrapping
# We want to run the circuit on JAX (for speed) but train in PyTorch (for usability)

with tc.runtime_backend("jax"):
    # 'torch_interface' wraps the JAX function into a PyTorch autograd function
    # jit=True enables XLA compilation on the JAX side
    q_func_torch = tc.interfaces.torch_interface(
        universal_circuit, jit=True, enable_dlpack=True
    )

# 3. PyTorch Training Loop (Host)
# The wrapped function behaves exactly like a native PyTorch operation
params_torch = torch.tensor([0.1, 0.1], requires_grad=True)

# Forward: PyTorch Tensor -> DLPack -> JAX (JIT Exec) -> DLPack -> PyTorch Tensor
loss = q_func_torch(params_torch)

# Backward: Gradients flow seamlessly back from JAX to PyTorch
loss.backward()

print(f"Loss: {loss.item():.4f}")
print(f"Gradients: {params_torch.grad}")
\end{lstlisting}

\subsection{Quantum Layers in Classical Pipelines}

\begin{mdframed}
\textbf{Reference Scripts: }
\fancylink{\rooturl examples/ng_whitepaper/IIIB_mnist_tf.py}{\sf IIIB\_mnist\_tf.py}, \fancylink{\rooturl examples/ng_whitepaper/IIIB_mnist_torch.py}{\sf IIIB\_mnist\_torch.py}
\end{mdframed}

Building upon the functional interoperability provided by backend interfaces, \tc\ provides high-level, object-oriented abstractions compatible with standard deep learning library definitions. We offer ready-to-use modules such as \texttt{tc.KerasLayer} and \texttt{tc.TorchLayer}, which encapsulate parameterized quantum circuits as standard trainable layers (e.g., Keras Layers or PyTorch Modules).
This design enables the seamless construction of hybrid quantum-classical neural networks, where quantum circuits act as plug-and-play components within classical architectures like ResNets or Transformers.

In the simplest regime, the quantum layer operates on the same backend as the classical model. The following example demonstrates constructing a Keras model where a tensor-native quantum circuit based data flow is inserted between classical dense layers to classify MNIST digits. The \texttt{tc.KerasLayer} automatically handles parameter initialization and backpropagation.

\begin{lstlisting}[language=Python]
# Backend: TensorFlow (Consistent with the classical model)
K = tc.set_backend("tensorflow")

def quantum_circuit(inputs, weights):
    c = tc.Circuit(n_qubits)
    # Encoding Layer
    for i in range(n_qubits):
        c.ry(i, theta=inputs[i])
    # Variational Layer (SU4 gates)
    for layer in range(n_layers):
        for i in range(n_qubits - 1):
            c.su4(i, i + 1, theta=weights[layer, i])
    # Measurement
    return K.stack([K.real(c.expectation_ps(z=[i])) for i in range(n_qubits)])

# Wrap as a Keras Layer
quantum_layer = tc.KerasLayer(
    quantum_circuit,
    weights_shape=[(n_layers, n_qubits - 1, 15)] # Shape of trainable parameters
)

# Construct Hybrid Model
model = tf.keras.Sequential([
    tf.keras.layers.Dense(n_qubits, input_shape=(784,)), # Classical Preprocessing
    quantum_layer,                                       # Quantum Processing
    tf.keras.layers.Dense(1, activation="sigmoid")       # Classical Postprocessing
])
\end{lstlisting}

Furthermore, these object-oriented layers can internally leverage the backend interfaces described in the previous subsection to achieve cross-framework acceleration.
A common high-performance pattern is to define the overall model in PyTorch (utilizing its rich ecosystem for data loading and classical layers) while configuring the internal \texttt{QuantumLayer} to execute on the JAX backend.
By setting \texttt{use\_interface=True}, \texttt{use\_jit=True}, and \texttt{enable\_dlpack=True}, the layer transparently handles the JIT compilation and zero-copy tensor transfer, providing JAX-level speed within a PyTorch workflow. This architectural flexibility allows \tc\ to combine the ecosystem usability of one framework with the raw computational performance of another.

\begin{lstlisting}[language=Python]
# Backend: JAX (For high-performance quantum simulation)
tc.set_backend("jax")

# Wrap as a PyTorch Module with internal JAX execution
quantum_layer = tc.TorchLayer(
    quantum_circuit,
    weights_shape=[n_layers, n_qubits - 1, 15],
    use_vmap=True,       # Auto-vectorize over the batch dimension
    use_interface=True,  # Enable cross-backend execution
    use_jit=True,        # JIT compile the JAX circuit
    enable_dlpack=True   # Zero-copy data transfer
)

class HybridModel(nn.Module):
    def __init__(self):
        super().__init__()
        self.fc_in = nn.Linear(784, n_qubits)
        self.quantum = quantum_layer # Acts like a standard nn.Module
        self.fc_out = nn.Linear(n_qubits, 1)

    def forward(self, x):
        x = self.fc_in(x)
        # PyTorch Tensor -> DLPack -> JAX (JIT/VMAP) -> DLPack -> PyTorch Tensor
        x = self.quantum(x) 
        x = torch.sigmoid(self.fc_out(x))
        return x
\end{lstlisting}

\subsection{Tensor Network Object Translation}
\begin{mdframed}
\textbf{Reference Script: }
\fancylink{\rooturl examples/ng_whitepaper/IIIC_tensornetwork_translation.py}{\sf IIIC\_tensornetwork\_translation.py}
\end{mdframed}

Recognizing the diverse ecosystem of tensor network tools, \tc\ implements translation utilities to convert tensor network like objects across different frameworks including \texttt{TensorNetwork}, \texttt{TeNPy} and \texttt{Quimb}. This interoperability allows researchers to design circuits in \tc, export them to specialized external contraction optimizers, and import the optimized path or result back into the differentiable pipeline.
\tc\ acts as a central hub, enabling a fluid workflow where users can leverage the best-in-class tool for each stage of a tensor network based simulation.

To achieve this, we provide native support for the \texttt{tensornetwork} \texttt{Node} structure and implement a bidirectional translation layer within the \texttt{tensorcircuit.quantum} module. The core abstraction is the \texttt{QuOperator} (and \texttt{QuVector}) restricted to matrix product state and matrix product operator formats.

This interoperability unlocks powerful hybrid workflows. For instance, a researcher can define a variational ansatz in \tc, export the state to \texttt{TeNPy} to perform canonicalization or apply time-evolving block decimation updates, and then import the resulting state back into \tc\ to calculate gradients or compute observables using GPU acceleration. 

\begin{lstlisting}[language=Python]
import tensorcircuit as tc
from tensorcircuit import quantum as qu
import tensornetwork as tn

# 1. Native Integration: Constructing from Raw Nodes
# \tc utilizes Google's TensorNetwork Nodes as building blocks
nodes = [
    tn.Node(data, name=f"site_{i}", axis_names=["vL", "p", "vR"]) 
    for i, data in enumerate(tensors)
]
# Create a TensorCircuit QuOperator (MPS wrapper)
tc_mps = qu.QuOperator(out_edges=[n[1] for n in nodes], in_edges=[])

# Bridge to TeNPy (e.g., for Canonical Form)
# Convert TC object to TeNPy MPS
tenpy_mps = qu.qop2tenpy(tc_mps)

# Perform physics-specific operations in TeNPy
tenpy_mps.canonical_form() 
print(f"TeNPy MPS Norm: {tenpy_mps.norm}")

# Convert back to TC for further execution
tc_mps_restored = qu.tenpy2qop(tenpy_mps)

# Or bridge to Quimb
quimb_tn = qu.qop2quimb(tc_mps)
tc_mps_restored = qu.quimb2qop(quimb_tn)

# Or bridge to TensorNetwork
tn_mps = qu.qop2tn(tc_mps)
tc_mps_restored = qu.tn2qop(tn_mps)
\end{lstlisting}

\subsection{Differentiable Circuit Object Transformation}

\begin{mdframed}
\textbf{Reference Script: }
\href{\rooturl examples/ng_whitepaper/IIID_qiskit_translation_vqe.py}{\sf IIID\_qiskit\_translation\_vqe.py}
\end{mdframed}

While \tc\ provides a concise native API for circuit construction, we recognize the immense value of the existing quantum software ecosystem. To facilitate interoperability, \tc\ supports standard circuit interchange formats, including bidirectional conversion for \textbf{OpenQASM} (\texttt{from\_openqasm, to\_openqasm}) and  bidirectional object translation to and from \textbf{Qiskit} and \textbf{Cirq}~\cite{cirq_developers_2025_16867504} circuit objects.

Beyond static circuit conversion, \tc\ enables a powerful dynamic translation workflow. This feature allows parameterized circuits defined in external libraries such as Qiskit to be directly embedded into the differentiable computational graph of TensorCircuit-NG.
Users can convert a Qiskit ansatz into a \tc\ function while dynamically binding JAX/TensorFlow tensors to the circuit parameters.

Crucially, when combined with JIT compilation, this translation incurs zero runtime overhead during the optimization loop. The translation logic is traced once, and the resulting tensor network contraction path is compiled into optimized machine code. This allows researchers to utilize Qiskit's rich library of heuristic ansatzes while leveraging the gradient-based optimization performance of \tc.

\begin{lstlisting}[language=Python]
import tensorcircuit as tc
import jax
import optax
from qiskit.circuit.library import RealAmplitudes

# 1. Leverage Qiskit's extensive library of pre-built Ansatzes
# Define an 8-qubit hardware-efficient ansatz
qiskit_ansatz = RealAmplitudes(num_qubits=8, reps=3, entanglement="linear")

def vqe_loss(params):
    # 2. Dynamic Translation with Parameter Binding
    # Instead of static conversion, we bind differentiable tensors params
    # directly to the Qiskit circuit structure during the forward pass.
    c = tc.translation.qiskit2tc(qiskit_ansatz, binding_params=params)
    
    # 3. Compute Expectation
    # The translated circuit c is fully compatible with tc's physics engine
    e = tc.templates.measurements.operator_expectation(c, hamiltonian)
    return tc.backend.real(e)

# 4. JIT Compilation
# The translation overhead is eliminated after the first trace.
# We get XLA-compiled performance for a Qiskit-defined circuit.
vqe_step = tc.backend.jit(tc.backend.value_and_grad(vqe_loss))

# Optimization Loop
params = jax.random.normal(key, shape=(qiskit_ansatz.num_parameters,))
optimizer = optax.adam(learning_rate=2e-2)

for i in range(300):
    loss, grads = vqe_step(params)
    updates, opt_state = optimizer.update(grads, opt_state)
    params = optax.apply_updates(params, updates)
\end{lstlisting}

\subsection{Domain-Specific Application: TenCirChem-NG}

\begin{mdframed}
\textbf{Reference Repository: }
\fancylink{https://github.com/tensorcircuit/tencirchem-ng}{\sf TenCirChem-NG}
\end{mdframed}

The versatility of \tc\ extends beyond generic circuit simulation; it serves as a robust foundation for building domain-specific quantum software. A prime example is \textbf{TenCirChem-NG}~\cite{Li2023tcc}, a specialized library designed for quantum chemistry~\cite{McArdle2020review, Cao2019review} and molecular simulation built on top of the \tc\ core.

TenCirChem-NG demonstrates how the unified programming paradigm enables specialized high-performance applications:

\begin{itemize}
    \item \textbf{Hybrid Simulation Engines:} TenCirChem-NG introduces a dual-engine architecture. It utilizes the standard \texttt{tensornetwork} engine of \tc\ for general parameterized circuits while implementing a specialized engine for unitary coupled cluster ansatzes~\cite{Anand2022}. This specialized engine exploits particle number conservation to store wavefunctions in configuration interaction vectors rather than the full Hilbert space, enabling the simulation of large molecular systems.

    \item \textbf{Variational Quantum Dynamics:} Leveraging the AD engine of \tc, TenCirChem-NG provides a dedicated module for simulating time-dependent quantum phenomena. It implements time-dependent variational principles to model complex processes.

    \item \textbf{Seamless Noise and Hardware Integration:} Inheriting the noise modeling capabilities of \tc, the library supports noisy circuit simulation via the \texttt{tensornetwork-noise} engine, allowing researchers to evaluate the impact of gate errors and finite measurement shots on variational energies. Furthermore, it provides a unified \texttt{qpu} engine that allows chemistry experiments to be deployed directly onto physical quantum processors without code modification via \tc\ infrastructure.
\end{itemize}

TenCirChem-NG exemplifies how \tc\ empowers domain experts to build vertically integrated solutions that bridge the gap between abstract quantum algorithms and concrete scientific discovery.

\section{Modeling the Physical World}

\subsection{Lattice and Hamiltonian}
\begin{mdframed}
\textbf{Reference Script: }
\href{\rooturl examples/ng_whitepaper/IVA_lattice_hamiltonian.py}{\sf IVA\_lattice\_hamiltonian.py}
\end{mdframed}

Quantum many-body physics often originates from geometrical arrangements of interacting particles. To support research in quantum many-body physics, \tc\ includes a robust module for lattice construction and Hamiltonian generation. Unlike standard graph libraries, the \tc\ lattice module is designed with differentiability at its core, allowing physical coordinates to be treated as trainable parameters.

The \texttt{tc.templates.lattice} module provides a high-level API to generate standard lattice geometries, including square, triangular, honeycomb, kagome lattices, etc. These objects manage the site coordinates, boundary conditions (open or periodic), and nearest-neighbor connectivity graphs automatically.

Furthermore, to model realistic experimental conditions—such as atom loss in optical tweezers or impurities in solid-state systems—users can easily create custom topologies. The \texttt{CustomizeLattice} class allows for the dynamic removal or addition of sites (defects) or the modification of interaction while preserving the underlying coordinate. To assist in verifying the geometric setup, the lattice objects provide a built-in visualization utility. The \texttt{show()} method leverages \texttt{matplotlib} to render the lattice graph, accurately plotting sites at their physical coordinates and drawing edges to represent nearest-neighbor connectivity. This immediate visual feedback is essential for debugging boundary conditions and confirming the correct placement of defects before launching computationally expensive simulations.

\begin{lstlisting}[language=Python]
# 1. Standard Kagome Lattice
# Automatically computes neighbor lists and coordinate positions
kagome = tc.templates.lattice.KagomeLattice(
    size=(2, 2), 
    pbc=False, # Open Boundary Conditions
    precompute_neighbors=1
)
# visualization
kagome.show()

# 2. Customizing for Defects
# Convert to a mutable custom lattice structure
custom_lat = tc.templates.lattice.CustomizeLattice.from_lattice(kagome)

# Simulate a physical defect by removing a specific site
hole_id = custom_lat.get_identifier(site_index=5)
custom_lat.remove_sites([hole_id])

# The connectivity graph is automatically updated
print(f"Remaining sites: {custom_lat.num_sites}")
\end{lstlisting}

The \texttt{tc.templates.hamiltonians} module enables the automated construction of operator lists from lattice objects. Because the entire pipeline—from coordinate definition to Hamiltonian construction—is end-to-end differentiable and jittable, researchers can optimize physical geometric parameters (such as the lattice constant) to minimize ground state energy or target specific spectral properties.

The following example demonstrates optimizing the lattice constant $a$ of a Rydberg Hamiltonian on triangular lattice toward a target ground state energy.

\begin{lstlisting}[language=Python]
# Physical Constants
omega, delta, c6 = 1.0, 1.0, 10.0

def get_ground_state_energy(a):
    # 1. Differentiable Lattice Construction
    # The lattice coordinates depend on the scalar input 'a'
    lat = tc.templates.lattice.TriangularLattice(
        size=(2, 4), 
        lattice_constant=tc.backend.abs(a), 
        pbc=False
    )

    # 2. Hamiltonian Generation
    # Interaction terms V_ij are computed from coordinates: C6 / |r_i - r_j|^6
    # Gradients propagate through the distance matrix calculation
    h_sparse = tc.templates.hamiltonians.rydberg_hamiltonian(
        lat, omega=omega, delta=delta, c6=c6
    )
    
    # 3. Energy Calculation (e.g., via dense diagonalization for small systems)
    h_dense = tc.backend.to_dense(h_sparse)
    evals = tc.backend.eigh(h_dense)[0]
    return evals[0] # Ground state energy

# Automatic Differentiation setup
# We compute the gradient of energy w.r.t the lattice constant 'a'
grad_fn = tc.backend.jit(tc.backend.value_and_grad(get_ground_state_energy))

# Optimization loop
a_param = tc.backend.convert_to_tensor(1.5) # Initial guess
optimizer = optax.adam(learning_rate=0.005)

for _ in range(100):
    energy, grads = grad_fn(a_param)
    updates, opt_state = optimizer.update(grads, opt_state)
    a_param = optax.apply_updates(a_param, updates)
\end{lstlisting}

This geometry-to-Hamiltonian end-to-end differentiability opens new avenues for hardware co-design, allowing algorithms to autonomously discover optimal qubit placements for specific quantum simulation tasks.

In many-body quantum simulation, constructing the sparse matrix representation of a Hamiltonian for large systems ($N > 20$) often becomes a computational bottleneck, especially when strictly relying on CPU-based serial execution. 
\tc\ resolves this bottleneck with the \texttt{PauliStringSum2COO} function, a fully vectorized, JIT-compiled implementation designed for GPU acceleration and the engine behind various lattice based Hamiltonian generation methods. Instead of row-by-row construction, this algorithm computes the coordinate list (COO) indices (row, col) and non-zero values (data) for all Pauli terms in parallel.
Crucially, this construction process is end-to-end differentiable and jittable as a traceable computational graph.

We benchmark the performance of constructing the sparse Hamiltonian matrix (TFIM) on an NVIDIA H200 GPU against standard CPU-based libraries (NumPy, QuSpin, Quimb). The results, detailed in Table~\ref{tab:hamiltonian_bench}, demonstrate that \tc\ on GPU achieves orders-of-magnitude speedups—constructing a $L=24$ sparse matrix in under 60 milliseconds, whereas traditional methods require tens of seconds.

\begin{table}[h]
\centering
\caption{Benchmark of sparse Hamiltonian construction time in complex128 precision: comparison performed on NVIDIA H200 (GPU) and Apple M4 Pro (CPU). \tc\ enables efficient construction of high dimensional Hamiltonians.}
\label{tab:hamiltonian_bench}
\begin{tabular}{@{}lllc@{}}
\toprule
\textbf{System Size~~} & \textbf{Library / Backend~} & \textbf{Device~} & \textbf{~Time (s)~} \\ \midrule
\textbf{22 Qubits} & \textbf{\tc\ (JAX)} & \textbf{GPU} & \textbf{0.019} \\
 & \tc\ (JAX) & CPU & 1.2 \\
 & \tc\ (NumPy) & CPU & 8.5 \\

 & QuSpin & CPU & 4.3 \\
 & Quimb & CPU & 10.9 \\ \midrule
\textbf{24 Qubits} & \textbf{\tc\ (JAX)} & \textbf{GPU} & \textbf{0.059} \\
 & \tc\ (JAX) & CPU & 14.1 \\
 &\tc\ (NumPy) & CPU & 50.4 \\
 & QuSpin & CPU & 20.4 \\
 & Quimb & CPU & 66.7 \\ \bottomrule
\end{tabular}
\end{table}

\subsection{Qudit Systems}

\begin{mdframed}
\textbf{Reference Script: }
\href{\rooturl examples/ng_whitepaper/IVB_qudit_simulation.py}{\sf IVB\_qudit\_simulation.py}
\end{mdframed}

While qubits ($d=2$) are the standard information carriers, high-dimensional qudit systems ($d \ge 3$) offer different advantages, including larger state spaces and efficient encoding of bosonic modes. However, software support for qudits has traditionally been fragmented or inefficient.
\tc\ treats the local Hilbert space dimension $d$ as a first-class citizen. The core engine is agnostic to tensor dimensions, allowing for the seamless simulation of qudits ($d>2$).

We introduce a dedicated \texttt{QuditCircuit} class that generalizes standard gate and measurement operations. For example, the parameterized rotation gates are extended to allow rotations within any two-level subspace spanned by levels $|j\rangle$ and $|k\rangle$.

To demonstrate this capability, we simulate the quantum clock model (a $\mathbb{Z}_d$ generalization of the Ising model) for a system of qutrits ($d=3$). The Hamiltonian is given by:
\begin{equation}
    H = - J \sum_i (Z_i Z_{i+1}^\dagger + \text{h.c.}) - h \sum_i (X_i + X_i^\dagger),
\end{equation}
where $Z$ and $X$ are the generalized clock and shift operators for qudits, respectively.

The following code snippet illustrates defining a hardware-efficient ansatz for qutrits and optimizing the ground state using the same JIT-compiled, batched VQE workflow commonly used for qubits. This qudit-native support ensures that researchers can explore quantum systems with high-dimensional local Hilbert space while utilizing the full performance of the differentiable and jittable engine.

\begin{lstlisting}[language=Python]
# 1. Define Qudit Circuit (d=3)
def ansatz(params, n=4, d=3):
    c = tc.QuditCircuit(n, dim=d)
    
    # Layered Ansatz
    for l in range(layers):
        # Local SU(d) rotations
        for i in range(n):
            # RZ gates on each level
            for k in range(d):
                c.rz(i, theta=params[l, i, k], j=k)
            # RY gates on subspaces (j, k)
            for j in range(d):
                for k in range(j+1, d):
                    c.ry(i, theta=params[l, i, idx], j=j, k=k)
        
        # Entanglement: Generalized C-SUM gate |x, y> -> |x, x+y mod d>
        for i in range(n-1):
            c.csum(i, i+1)
            
    return c

# 2. Hamiltonian expectation
def get_energy(params):
    c = ansatz_circuit(params)
    return tc.templates.measurements.operator_expectation(c, H_sparse)


# 3. Batch Optimization
# We leverage vmap to optimize multiple random initializations in parallel
# The optimizer and gradient updates are identical to the qubit case
vqe_step = tc.backend.jit(tc.backend.vmap(tc.backend.value_and_grad(get_energy)))
\end{lstlisting}

\subsection{Fermion Gaussian States}

\begin{mdframed}
\textbf{Reference Scripts: }
\href{\rooturl examples/ng_whitepaper/IVC_fgs_groundstate.py}{\sf IVC\_fgs\_groundstate.py}, \href{\rooturl examples/ng_whitepaper/IVC_fgs_dynamics.py}{\sf IVC\_fgs\_dynamics.py}
\end{mdframed}

While general quantum many-body simulation scales exponentially with system size, a significant class of physically relevant systems—non-interacting fermions~\cite{Terhal2002} also related to matchgate circuits~\cite{Valiant2002}—can be simulated efficiently in polynomial time. \tc\ includes a dedicated module for fermion Gaussian states (FGS)~\cite{Surace2022, Ravindranath2025}, allowing researchers to simulate systems with thousands of fermions using the covariance matrix formalism.

Unlike the general tensor network engine which tracks the full $2^N$ state vector or an MPS approximation, the FGS module tracks the two-body correlation matrix $\Gamma_{ij} = \langle c_i^\dagger c_j \rangle$ including pairing correlations $\langle c_i c_j \rangle$ for systems without particle conservation~\cite{Guo2025}. This reduces the computational complexity of ground state preparation and time evolution to $O(N^3)$, while retaining the ability to compute multi-point correlations and entanglement entropy related quantities exactly.

The module provides automated routines to compute the ground state of any quadratic Hamiltonian $H = \sum_{ij} A_{ij} c_i^\dagger c_j + \frac{1}{2} \sum_{ij} (B_{ij} c_i^\dagger c_j^\dagger + \text{h.c.})$. We demonstrate the usage by studying topological phase transitions in the 1D Kitaev chain~\cite{Kitaev2001}.
Crucially, the entire FGS pipeline in \tc\ is differentiable. Users can compute the gradient of macroscopic properties (like entanglement entropy) with respect to Hamiltonian parameters, enabling the automated discovery of phase boundaries.
 By maximizing the entanglement entropy via gradient ascent, the solver locates the critical chemical potential $\mu_c = 2t$ where the Majorana zero modes emerge. The differentiability of the FGS simulator thus allows for Hamiltonian learning and inverse engineering.

\begin{lstlisting}[language=Python]
# 1. Define Hamiltonian Construction (Differentiable w.r.t parameters)
def get_entropy(mu, L=200):
    # Construct Kitaev Hamiltonian matrix (2L x 2L)
    # H = -t sum(c+ c) + Delta sum(c c) - mu sum(n)
    h_mat = get_kitaev_hamiltonian(L, t=1.0, delta=1.0, mu=mu)
    
    # 2. Solve for Ground State (Differentiable Eigensolver)
    # The FGS simulator diagonalizes the matrix to find the correlation matrix
    sim = tc.FGSSimulator(L, hc=h_mat)
    
    # 3. Compute Half-Chain Entanglement Entropy
    return tc.backend.real(sim.entropy(subsystems_to_trace_out=range(L//2, L)))

# 4. Automated Phase Detection
# Compute gradient of Entropy w.r.t Chemical Potential
grad_fn = tc.backend.jit(tc.backend.value_and_grad(get_entropy))

# Optimization loop automatically finds the critical point where entropy peaks
mu_guess = 1.0
loss, grads = grad_fn(mu_guess)
\end{lstlisting}

Beyond static properties, \tc\ supports efficient time evolution for both real-time (quenches) and imaginary-time non-unitary (cooling) dynamics~\cite{Ravindranath2023, Chang2024} for FGS.
This capability enables the simulation of quantum quenches in large systems inaccessible to exact diagonalization. The code below simulates the melting of a N\'eel state order and the resulting light-cone spreading of correlations.

\begin{lstlisting}[language=Python]
# Setup: 60-site chain initialized in Neel State |1010...>
L = 60
sim = tc.FGSSimulator(L, filled=[2*i for i in range(L//2)])

# Quench Hamiltonian (Pure Hopping)
h_hopping = get_hopping_hamiltonian(L, t=1.0)
time_step_op = h_hopping * dt * 2
# Note evol_hamiltoian implements exp(-iHt/2)

correlations = []
for step in range(100):
    # 1. Compute Time-Separated Correlation (OTOC-style)
    # Measure correlation between site i at time t and center site at t=0
    # <c_i^dag(t) c_{center}(0)>
    cm_otoc = sim.get_cmatrix(now_i=True, now_j=False)
    correlations.append(jnp.abs(cm_otoc[L:, L//2 + L])**2)
    
    # 2. Evolve State
    sim.evol_hamiltonian(time_step_op)

# The result visualizes the linear light-cone propagation of information
\end{lstlisting}

Beyond standard observables, the FGS module includes specialized routines for characterizing complex quantum phenomena related to symmetry and measurement monitored dynamics such as efficient implementations for calculating entanglement asymmetry~\cite{Ares2023ea} and subsystem information capacity~\cite{Chen2025sic, Qing2025}.
Understanding how entanglement intertwines with internal symmetries is crucial for classifying topological phases and studying non-equilibrium symmetry breaking (e.g., the quantum Mpemba effect~\cite{Ares2025review, Yu2025review}). Moreover, the interplay between unitary evolution and projective measurements leads to rich phenomena such as measurement-induced phase transitions (MIPT) in fermionic systems~\cite{Cao2019fmipt, Chen2020fmipt}. The FGS simulator supports efficient conditional measurements and post-selection on specific fermionic sites. Unlike full state-vector simulations, the measurement update in FGS is implemented via low-rank updates to the correlation matrix, maintaining polynomial complexity for hybrid fermionic circuits with frequent mid-circuit measurements.

\subsection{Time Evolution}

\begin{mdframed}
\textbf{Reference Script: }
\href{\rooturl examples/ng_whitepaper/IVD_time_evolution.py}{\sf IVD\_time\_evolution.py}
\end{mdframed}

Modeling the dynamics of quantum many-body systems is a cornerstone of condensed matter physics~\cite{Eisert2015, Yu2025}. \tc\ provides a comprehensive suite of time evolution solvers, ranging from exact methods for small systems to advanced approximation techniques for larger Hilbert spaces. These solvers are built on the backend's sparse linear algebra primitives, enabling hardware acceleration and automatic differentiation through the time evolution operator $U(t)$.

For time-independent Hamiltonians $H$, the evolution $|\psi(t)\rangle = e^{-iHt}|\psi(0)\rangle$ can be computed using several strategies depending on the system size and spectrum:

\begin{itemize}
    \item \textbf{Exact Diagonalization (ED):} Computes the full matrix exponential. Best for small systems ($N \lesssim 14$) where dense matrix operations are feasible.
    \item \textbf{Krylov Subspace Methods (Lanczos):} Approximates the evolution in a small subspace spanned by $\{|\psi\rangle, H|\psi\rangle, \dots, H^k|\psi\rangle\}$. This method is highly efficient for intermediate systems ($N \sim 20-30$) as it only requires sparse matrix-vector multiplication.
    \item \textbf{Chebyshev Expansion:} Expands the evolution operator in terms of Chebyshev polynomials. This method offers uniform convergence error and is particularly robust for long-time dynamics, provided the spectral bounds of $H$ are known.
\end{itemize}

The following snippet compares these methods for a 1D Heisenberg chain under a random magnetic field.

\begin{lstlisting}[language=Python]
# 1. Exact Diagonalization

ed_evol(h_dense, psi0, 1j * times)


# 2. Krylov Subspace Evolution (Lanczos)
# Efficient for sparse Hamiltonians, computes exp(-iHt) implicitly
# subspace_dimension controls the size of the Krylov basis
states_krylov = tc.timeevol.krylov_evol(
    h_sparse, psi0, times, subspace_dimension=60
)

# 3. Chebyshev Expansion
# Requires spectral bounds [e_min, e_max] to normalize the Hamiltonian
e_max, e_min = tc.timeevol.estimate_spectral_bounds(h_sparse)
k, M = tc.timeevol.estimate_k(t_max, (e_max, e_min)) # Estimate expansion order

state_chebyshev = tc.timeevol.chebyshev_evol(
        h_sparse, psi0, t, (e_max, e_min), k, M
    )
\end{lstlisting}

For systems driven by time-dependent Hamiltonian $H(t)$, analytical solutions are rarely available. \tc\ integrates with ordinary differential equation (ODE) solvers (e.g., \texttt{jax.experimental.ode} or \texttt{diffrax}) to numerically integrate the Schrödinger equation $i\frac{d}{dt}|\psi(t)\rangle = H(t)|\psi(t)\rangle$ with AD support~\cite{Chen2018ode}.
This approach allows for the simulation and design of arbitrary driving protocols, such as Floquet engineering or adiabatic passage.

\begin{lstlisting}[language=Python]
# Define time-dependent Hamiltonian: H(t) = H_static + A * cos(omega * t) * X_0
def h_driven(t, A, omega):
    driving_term = A * tc.backend.cos(omega * t) * x0_op
    return h_static + driving_term

# Integrate Schrodinger equation
# The solver automatically handles the time-dependent generator
states_driven = tc.timeevol.ode_evol_global(
    h_driven,
    psi0,
    times,
    solver_args=(1.0, 2.0), # A=1.0, omega=2.0
    ode_backend="jaxode",
    rtol=1e-6
)
\end{lstlisting}

\subsection{Noise Profile: Configuration, Simulation, and Mitigation}
\begin{mdframed}
\textbf{Reference Script: }
\href{\rooturl examples/ng_whitepaper/IVE_noise_customization.py}{\sf IVE\_noise\_customization.py}
\end{mdframed}

In the original TensorCircuit, we established the fundamental capacity for open quantum systems, supporting both exact density matrix evolution and scalable Monte Carlo trajectories simulation.
Building upon this robust computational backbone, \tc\ now advances the modeling capability by introducing a high-level noise profile customization engine.
This upgrade shifts the focus from merely executing noisy circuits to faithfully modeling physical hardware. By enabling rich noise configurations, we empower researchers to construct heterogeneous error environments that mirror the specific imperfections of real-world experimental setups, rather than relying on idealized, uniform noise models.
Users can attach specific quantum channels (e.g., amplitude damping, thermal relaxation, depolarizing) to gates based on complex criteria, such as qubit indices, gate types, or even custom logic functions.
This flexibility is crucial for modeling heterogeneous hardware where coherence times ($T_1, T_2$) vary across the chip.

\begin{lstlisting}[language=Python]
from tensorcircuit.noisemodel import NoiseConf

# 1. Global Noise: Depolarizing on all CNOTs
noise_conf = NoiseConf()
noise_conf.add_noise("cnot", tc.channels.generaldepolarizingchannel(1e-3, 2))

# 2. Site-Dependent Noise
# Simulating a device where higher indexed qubits are noisier
for i in range(n_qubits):
    rate = 0.005 * (i + 1)
    noise_conf.add_noise("ry", [tc.channels.phasedampingchannel(rate)], qubit=[(i,)])

# 3. Logic-Based Noise: Apply only to specific topology
def edge_condition(d):
    # Apply noise only if the gate acts on even qubits
    return d["gatef"].n == "ry" and d["index"][0] % 2 == 0

noise_conf.add_noise_by_condition(
    edge_condition, tc.channels.amplitudedampingchannel(0.05)
)
\end{lstlisting}

Readout errors are often the dominant noise source in near-term devices. \tc\ integrates a dedicated workflow for simulating classical readout confusion matrices and mitigating them both in a scalable way. The \texttt{ReadoutMit} module automates the readout error calibration and mitigation process, closing the loop on realistic hardware simulation.

\begin{lstlisting}[language=Python]
# 1. Inject Readout Error
# Define error rates: p(0|0)=0.95, p(1|1)=0.92
readout_params = [[0.95, 0.92]] * n_qubits
noise_conf.add_noise("readout", readout_params)

# 2. Simulation (Virtual Hardware)
# Generate noisy counts from the circuit
c_noisy = tc.noisemodel.circuit_with_noise(c, noise_conf)
raw_counts = c_noisy.sample(batch=8192, readout_error=noise_conf.readout_error)

# 3. Mitigation
# Initialize mitigator with the execution interface
mitigator = tc.results.readout_mitigation.ReadoutMit(execute_fn)

# Run calibration circuits and invert the confusion matrix
mitigator.cals_from_system(qubit_list=range(n_qubits))
mitigated_counts = mitigator.apply_correction(raw_counts)
exp_val_mitigated = tc.results.counts.expectation(mitigated_counts, z=[n // 2])
\end{lstlisting}

\section{Advanced Simulation Engines}

\subsection{Analog Simulator}
\begin{mdframed}
\textbf{Reference Script: }
\href{\rooturl examples/ng_whitepaper/VA_analog_circuit.py}{\sf VA\_analog\_circuit.py}
\end{mdframed}

While standard quantum compilers abstract physics into discrete gate sequences, real quantum hardware operates via continuous-time control. To model these dynamics without the approximation errors and depth overhead of Trotterization~\cite{Lloyd1996}, \tc\ introduces the \texttt{AnalogCircuit} interface for analog quantum simulation.
This module treats the time evolution operator $U(t) = \mathcal{T} e^{-i \int_0^t H(\tau) d\tau}$ as a native differentiable primitive. By integrating with neural ODE solvers, \tc\ allows for the simulation of arbitrary time-dependent Hamiltonians with high precision.

A primary application of this capability is digital-analog hybrid quantum computing~\cite{Parra-Rodriguez2020}. In this paradigm, circuits are constructed by interleaving digital gates with global or local multi-qubit analog evolution blocks.
The following example demonstrates a digital-analog hybrid VQE for the TFIM. A unique feature of \tc's architecture is that the time duration of the analog blocks can also be treated as a trainable parameter. The optimizer learns not just the gate parameters, but also the optimal physical interaction time and strength required to reach the target state.
This approach allows researchers to explore hardware-efficient ansatzes that utilize the native connectivity and interaction of the device, such as superconducting processors or Rydberg atom arrays~\cite{Browaeys2020, Liu2025}.

\begin{lstlisting}[language=Python]
# Define the continuous Hamiltonian terms
# H_int = ZZ interaction, H_drive = Transverse X field
def h_interaction(t): return H_int_dense
def h_drive(t): return H_drive_dense

def hybrid_ansatz(params):
    # params contains both digital angles and analog durations
    gate_params, time_params = params
    
    # Initialize AnalogCircuit (supports ODE integration)
    c = tc.AnalogCircuit(nqubits)
    c.set_solver_options(ode_backend="diffrax")

    for d in range(depth):
        # 1. Digital Block: Single-qubit control
        for i in range(nqubits):
            c.rx(i, theta=gate_params[d, i, 0])
            c.ry(i, theta=gate_params[d, i, 1])

        # 2. Analog Block: Global Hamiltonian Evolution
        # Crucially, the evolution time t is differentiable
        t_int = tc.backend.abs(time_params[d, 0])
        c.add_analog_block(h_interaction, time=t_int)
        
        t_drive = tc.backend.abs(time_params[d, 1])
        c.add_analog_block(h_drive, time=t_drive)

    return c
\end{lstlisting}

\subsection{Stabilizer Simulator}
\begin{mdframed}
\textbf{Reference Script: }
\href{\rooturl examples/ng_whitepaper/VB_stabilizer_mipt.py}{\sf VB\_stabilizer\_mipt.py}
\end{mdframed}

For circuits consisting exclusively of Clifford gates and Pauli measurements, the Gottesman-Knill theorem allows for efficient simulation with polynomial complexity~\cite{Gottesman1998}. \tc\ integrates \textbf{Stim}~\cite{Gidney2021}, the state-of-the-art stabilizer simulator, into its \texttt{StabilizerCircuit} interface. This enables researchers to simulate systems with thousands of qubits using the same familiar API used for standard circuit construction in \tc\, which is essential for studying quantum error correction and entanglement phase transitions.

A paradigmatic application of this engine is the study of MIPT in Clifford circuits~\cite{Li2018mipt, Li2019b, Skinner2019a, Chan2019}. This phenomenon arises from the competition between random unitary dynamics~\cite{Fisher2023, Liu2025review} and projective measurements.
To resolve the critical measurement probability $p_c$ where this transition occurs, one must simulate system sizes far beyond the reach of state-vector methods. The \texttt{StabilizerCircuit} allows for the rapid simulation of multiple quantum trajectories to calculate the steady-state entanglement entropy for large-scale circuits. By sweeping the measurement probability $p$ across different system sizes $L$, users can observe the crossing point in the entanglement entropy, identifying the critical threshold separating the volume-law and area-law phases.

\begin{lstlisting}[language=Python]
def run_trajectory(L, p, depth):
    # Initialize a stabilizer simulator backed by Stim
    c = tc.StabilizerCircuit(L)
    
    for t in range(depth):
        # 1. Scrambling: Random 2-qubit Clifford gates
        # The engine efficiently updates the stabilizer tableau
        for i in range(0, L, 2):
             c.random_gate(i, (i + 1) % L)
             
        # 2. Monitoring: Projective Measurements
        # Measure each site with probability p
        for i in range(L):
            if np.random.random() < p:
                c.cond_measure(i)
                
        # 3. Entanglement Entropy
        # In the stabilizer formalism, entropy is computed via
        # the rank of the clipped tableau (O(N^2) complexity)
        sent = c.entanglement_entropy(subsystem=list(range(L//2)))
            
    return sent
\end{lstlisting}

\subsection{Approximate Matrix Product State Simulator}
\begin{mdframed}
\textbf{Reference Script: }
\href{\rooturl examples/ng_whitepaper/VC_mps_vqe.py}{\sf VC\_mps\_vqe.py}
\end{mdframed}

For systems with limited entanglement, such as 1D chains or shallow circuits, MPS offer a highly efficient simulation paradigm~\cite{Schollwock2011, Zhou2020}. \tc\ implements a fully differentiable MPS simulator, \texttt{MPSCircuit}, which manages bond dimensions via singular value decomposition (SVD) truncation.

Unlike standard tensor network libraries that are often static, \tc's MPS module is deeply integrated with the automatic differentiation engine. This allows gradients to flow through the iterative complex-valued SVD truncation steps~\cite{Wan2019}, enabling variational optimization of large-scale approximate circuit simulation. It is important to note that this method is inherently approximate. By truncating the singular value spectrum, users explicitly trade off numerical precision for the ability to scale to significantly larger system sizes inaccessible to exact state-vector methods.

The following example demonstrates a VQE for a TFIM where the MPS approach can scale to hundreds of qubits by capping the maximum singular values kept during simulation. \texttt{MPSCircuit} supports the same set of API as \texttt{Circuit} class. Therefore, by seamlessly switching between the \texttt{Circuit} and \texttt{MPSCircuit} interfaces, researchers can benchmark exact results against MPS approximations with minimal code changes.

\begin{lstlisting}[language=Python]
def tfim_energy_mps(params):
    # Initialize MPS Circuit with truncation rules
    # max_singular_values controls the bond dimension (accuracy vs. speed)
    c = tc.MPSCircuit(n, split={"max_singular_values": 16})

    # --- Ansatz Construction ---
    k = 0
    # Entangling layers
    for i in range(0, n - 1, 2):
        c.rzz(i, i + 1, theta=params[k])
        k += 1
    # Single-qubit rotations
    for i in range(n):
        c.rx(i, theta=params[k])
        k += 1
        
    # --- Expectation Calculation ---
    e_total = 0.0
    # Interaction terms <Z_i Z_{i+1}>
    for i in range(n - 1):
        # Efficient local contraction
        val = c.expectation((tc.gates.z(), [i]), (tc.gates.z(), [i + 1]))
        e_total -= val
        
    # Transverse field terms <X_i>
    for i in range(n):
        val = c.expectation((tc.gates.x(), [i]))
        e_total -= g * val

    return tc.backend.real(e_total)

# Automatic Differentiation & JIT Compilation
# Gradients flow through the iterative SVD truncation steps
vqe_val_and_grad = K.jit(K.value_and_grad(tfim_energy_mps))

# Execute optimization step
energy, grads = vqe_val_and_grad(params)
\end{lstlisting}

\section{Distributed Computing Infrastructure}

\subsection{Data Parallelism: Scaling to Multiple Devices}

\begin{mdframed}
\textbf{Reference Scripts: }
\href{\rooturl examples/ng_whitepaper/VIA_pmap_vqe.py}{\sf VIA\_pmap\_vqe.py}, 
\href{\rooturl examples/ng_whitepaper/VIA_pmap_qml.py}{\sf VIA\_pmap\_qml.py}, 
\href{\rooturl examples/ng_whitepaper/VIA_sharding_vqe.py}{\sf VIA\_sharding\_vqe.py}, 
\href{\rooturl examples/ng_whitepaper/VIA_sharding_qml.py}{\sf VIA\_sharding\_qml.py}
\end{mdframed}

Scaling quantum simulations often requires distributing workloads across multiple devices. \tc's JAX backend inherits the powerful distributed computing capabilities of the XLA compiler, offering two programming paradigms for data parallelism: explicit SPMD (Single Program Multiple Data) via \texttt{jax.pmap} and automated parallelism via \texttt{jax.jit}.

These techniques address several key bottlenecks in simulation quantum algorithms and quantum dynamics:
\begin{enumerate}
    \item \textbf{Measurement Parallelism:} Distributing the evaluation of massive Hamiltonians ($H = \sum_{i=1}^M c_i P_i$) where the number of terms $M$ is large.
    \item \textbf{Batch Parallelism:} Distributing large datasets across devices to compute batched gradients for hybrid training.
    \item \textbf{Parameter Parallelism:} Simultaneously executing multiple optimization by batching different circuit parameters.
    \item \textbf{Trajectory Parallelism:} Distributing independent Monte Carlo trajectories for noisy simulations (e.g., quantum jumps or MIPT), where each device simulates a unique stochastic realization.
\end{enumerate}

\textbf{Paradigm 1: Explicit Parallelism with \texttt{jax.pmap}}.
The \texttt{jax.pmap} transform enables explicit parallelization. In this paradigm, the user manually shards input data (e.g., reshaping a batch of size $B$ into shape $[N_{dev}, B/N_{dev}, \dots]$). The same function is executed on every device, and gradients are manually synchronized using global reduction primitives like \texttt{jax.lax.pmean}.

\begin{lstlisting}[language=Python]
# 1. Define the Update Step (Runs locally on each device)
def update_step(params, opt_state, x_batch, y_batch):
    # Compute local gradients
    loss, grads = jax.value_and_grad(loss_fn)(params, x_batch, y_batch)
    
    # 2. Explicit Synchronization
    # Average gradients across all devices "i"
    grads = jax.lax.pmean(grads, axis_name="i")
    
    updates, opt_state = optimizer.update(grads, opt_state)
    return optax.apply_updates(params, updates), opt_state

# 3. Parallelize the function
# axis_name='i' creates the communication group
parallel_update = jax.pmap(update_step, axis_name="i")

# 4. Execution
# Input data must be reshaped to (n_devices, batch_per_device, ...)
params_replicated = jax.device_put_replicated(params, devices)
parallel_update(params_replicated, opt_state, x_sharded, y_sharded)
\end{lstlisting}

\textbf{Paradigm 2: Automated Parallelism with \texttt{jax.jit} and \texttt{jax.sharding}}.
The second, more modern approach leverages JAX's SPMD via \texttt{jax.jit}. In this paradigm, the user writes code as if it were running on a single global device. By defining a logical device \texttt{Mesh} and annotating tensor layouts using \texttt{NamedSharding}, the XLA compiler automatically partitions the computation and inserts necessary communication protocols.
This approach significantly simplifies the code implementation, as it decouples the physics logic from the distribution logic.

\begin{lstlisting}[language=Python]
# 1. Define Device Mesh and Sharding Strategy
mesh = Mesh(jax.local_devices(), axis_names=("terms",))
# Shard the Hamiltonian terms along the "terms" axis
sharding = NamedSharding(mesh, PartitionSpec("terms"))

# 2. Distribute Data (Logical Global View)
# The compiler handles the physical distribution
weights_sharded = jax.device_put(weights, sharding)
structures_sharded = jax.device_put(structures, sharding)

# 3. Define Global Computation (Single-Device Style)
def loss_fn(params, w, s):
    # vmap over terms
    # JAX detects w and s are sharded and automatically
    # distributes this vmap loop across the mesh
    expt = jax.vmap(term_expectation)(w, s)
    return jnp.sum(expt)

# 4. JIT Compile
# The compiler generates the distributed program automatically
@jax.jit
def update(params, opt_state, w, s):
    loss, grads = jax.value_and_grad(loss_fn)(params, w, s)
    updates, opt_state = optimizer.update(grads, opt_state)
    return optax.apply_updates(params, updates), opt_state
\end{lstlisting}

Both paradigms allow \tc\ to scale linearly with the number of available accelerators, making it possible to train QML models on large datasets or simulate Hamiltonians with millions of terms.

\subsection{Model Parallelism: Distributed Tensor Network Contraction}
\begin{mdframed}
\textbf{Reference Module: }
\href{\rooturl examples/multi_host}{\sf multi\_host}
\end{mdframed}

For quantum circuits that exceed the memory capacity of a single GPU, data parallelism is insufficient. In these cases, \tc\ employs model parallelism, where the tensor network contraction task itself is sliced and distributed across a mesh of devices or even across multi-node clusters.

We introduce the \texttt{DistributedContractor}, a module that integrates the \texttt{cotengra} hyper-optimizer for tensor network slicing and contraction path finding with JAX's distributed strategy. This enables the simulation of massive tensor networks by treating the entire device cluster as a single logical accelerator.

\textbf{The \texttt{DistributedContractor} Engine}.
At the heart of this capability is the \texttt{DistributedContractor} class. This high-level orchestrator abstracts away the complexity of slicing the tensor network and  mapping a tensor network contraction onto a multi-device mesh. It serves as a bridge between the physical definition of the circuit and the distributed JAX runtime.

Unlike standard tensor network libraries that operate on static lists of nodes, \texttt{DistributedContractor} requires a generative function, \texttt{nodes\_fn(params)}.
To support AD in a variational setting, JAX must trace the graph construction process at every step. The contractor calls this function internally to regenerate the computational graph with updated parameter values while strictly preserving the optimal contraction path found during initialization.
The core enabler of model parallelism is slicing—cutting several large contraction indices to decompose a massive tensor operation into a sum of smaller, memory-efficient chunks. 

Constructing the uncontracted tensor network manually (e.g., explicitly coding \texttt{psi.adjoint() @ op @ psi}) can be tedious and error-prone for complex observables. \tc\ provides two efficient ways to streamline the definition of \texttt{nodes\_fn}:
(1) {Built-in Helpers:} For standard tasks, methods like \texttt{c.expectation\_before()} and \texttt{c.amplitude\_before()} directly return the list of nodes without performing the final real contraction.
(2) {Runtime Capture:} For arbitrary high-level operations, \tc\ offers the \texttt{tc.cons.runtime\_nodes\_capture()} context manager. This tool intercepts the contraction call within any standard API, suspends execution, and extracts the underlying tensor graph.
This allows users to define the distributed physics logic using the familiar, high-level syntax they use for circuit simulations.

\begin{lstlisting}[language=Python]
def nodes_fn(params):
    c = circuit_ansatz(N, D, params)
    
    # "Intercept" the tensor network generation
    with tc.cons.runtime_nodes_capture() as captured:
        # Call standard high-level API
        # The context manager stops actual contraction and captures the graph
        c.expectation([tc.gates.z(), [0]])
    
    # Return the raw nodes to be distributed
    return captured.get("nodes")
\end{lstlisting}

The \texttt{DistributedContractor} also exposes slicing and contraction settings via \texttt{cotengra\_options}. Users can specify \texttt{target\_size} (maximum memory per slice) and other hyperparameters, allowing the contraction of circuits that are physically larger than the aggregate memory of the cluster.

\textit{Time-Space Trade-off on Single Device}:
Crucially, the utility of \texttt{DistributedContractor} extends beyond multi-device clusters. Even on a single GPU, it enables the simulation of circuits that far exceed physical memory.
By leveraging the slicing mechanism, the engine automatically manages the serial execution of tensor slices—trading computation time for memory space. This allows researchers to simulate high-width or high-depth circuits on a single workstation that would otherwise trigger Out-Of-Memory errors in standard simulators.

\textbf{Workflow: The Find-Execute Pattern}.
For large-scale cluster simulations, we recommend a two-stage workflow to decouple the expensive contraction path search from the distributed execution.

\textit{Stage 1: Offline Pathfinding}:
First, we run a pathfinding script on a single high-memory node (CPU). This step uses \texttt{cotengra} to identify optimal slices in the tensor network that minimize space and time complexity. The resulting contraction tree is saved to disk.

\begin{lstlisting}[language=Python]
# Define the dynamic physics generator
def nodes_fn(params):
    # Reconstruct the ansatz with new params
    psi = circuit_ansatz(N, D, params).get_quvector()
    # Return the uncontracted nodes of <psi|H|psi>
    return (psi.adjoint() @ mpo @ psi).nodes

# Run heavy-duty path optimizer
DistributedContractor.find_path(
    nodes_fn=nodes_fn,
    params=dummy_params,
    cotengra_options={
        "slicing_reconf_opts": {"target_size": 2**28}, # Max number of elements in one slice
        "max_repeats": 128, # Optimization trials
        "parallel": 4, # Number of processes for path finding
    },
    filepath="tree.pkl"  # Save optimal path for the cluster run
)
\end{lstlisting}

\textit{Stage 2: Distributed Execution}.
Next, we launch the distributed VQE training on the GPU cluster. The script initializes a JAX distributed environment, loads the pre-computed path, and creates a global device mesh. The \texttt{DistributedContractor} then executes the contraction in parallel, automatically handling peer-to-peer communication between GPUs.

\begin{lstlisting}[language=Python]
def run_vqe_main():
    # 1. Initialize Distributed Environment (Slurm/JAX)
    jax.distributed.initialize()
    global_mesh = Mesh(jax.devices(), axis_names=("devices",))

    # 2. Load Pre-computed Contraction Path
    # This avoids re-running the expensive path search on the GPU cluster
    DC = DistributedContractor.from_path(
        filepath="tree.pkl",
        nodes_fn=nodes_fn,
        mesh=global_mesh,
        params=params_sharded, # Params distributed across GPUs
    )

    # 3. Differentiable Distributed Optimization
    # The contraction logic is JIT-compiled and distributed
    loss, grads = DC.value_and_grad(params)
\end{lstlisting}

\textbf{HPC Cluster Deployment}.
This framework is designed for standard HPC schedulers like Slurm. The following launch script demonstrates how to coordinate multiple GPU nodes, automatically setting up the master address and port required for JAX's multi-host initialization.

\begin{lstlisting}[language=bash]
#SBATCH --nodes=2
#SBATCH --ntasks-per-node=1
#SBATCH --gres=gpu:4

# Auto-detect master node for JAX coordination
export MASTER_NODE=$(scontrol show hostnames $SLURM_JOB_NODELIST | head -n 1)
export MASTER_ADDR=$(srun --nodes=1 --ntasks=1 -w "$MASTER_NODE" hostname -I)
export MASTER_PORT=29500

# Launch JAX script on all allocated nodes
# Each process will self-identify and join the mesh
srun python slurm_vqe_with_path.py
\end{lstlisting}

By combining advanced tensor slicing with JAX's SPMD runtime, \tc\ enables the exact simulation of quantum circuits, specifically VQAs, at previously inaccessible scales.

\textbf{Benchmark}.
We evaluated the distributed VQE pipeline on a high-performance compute node equipped with 8 NVIDIA H200 GPUs (141GB memory per device), interconnected via NVLink to minimize communication latency. The benchmark task involves computing the gradients for a 1D TFIM. The circuit depth was scaled linearly with system size ($L=N/2$). Each layer of the circuit comprises two-qubit SU(4) gate in ladder layout, yielding $15L(N-1)$ trainable parameters for the given variational circuit.
We utilized \texttt{cotengra} with \texttt{max\_repeats=640} during the offline pathfinding phase with optimization target $\text{FLOPS}+640\times \text{WRITE}$.
To fit the massive intermediate tensors of the 40-qubit circuit into the memory, we set the slicing \texttt{target\_size} to $2^{29}$ elements. 

\begin{figure}[t] 
    \centering
    \includegraphics[width=0.98\textwidth]{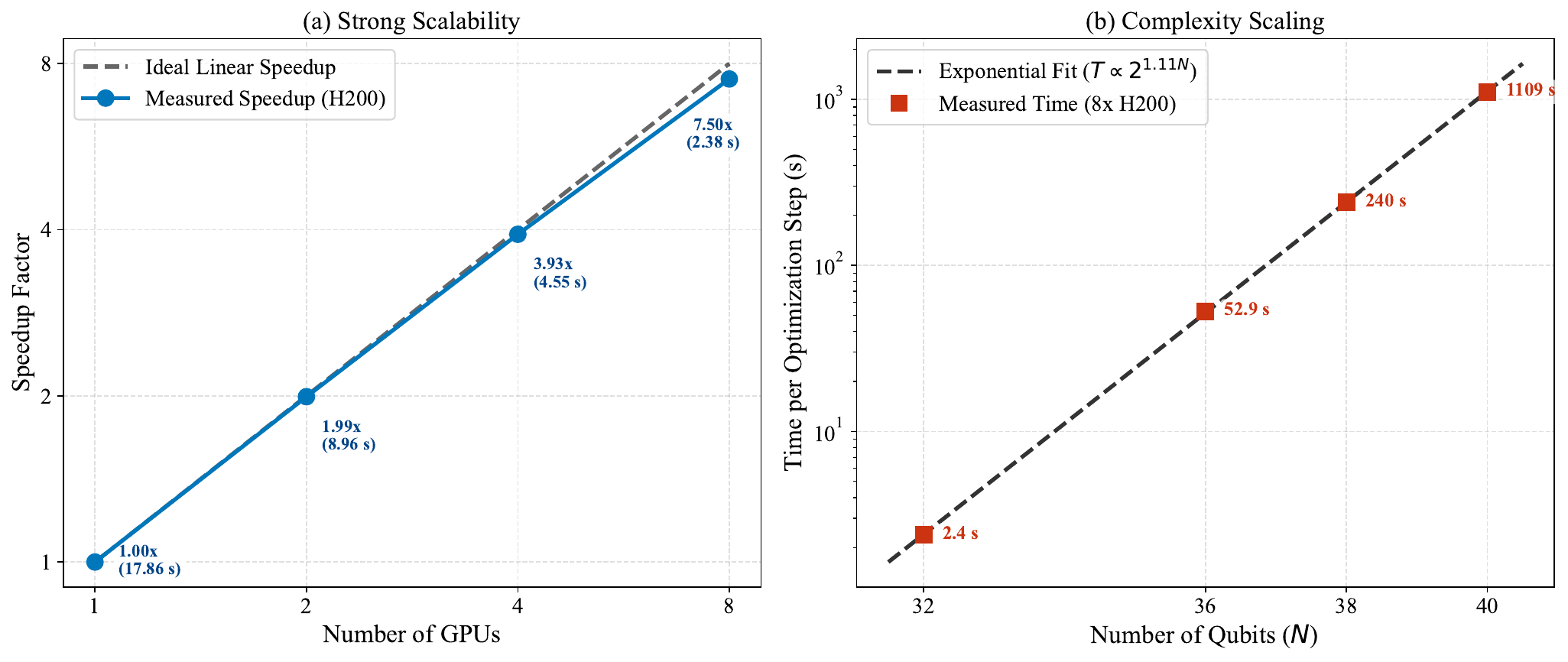}
    \caption{\textbf{Performance Benchmarks on Distributed VQE on NVIDIA H200 Cluster.} 
    (a) Strong scalability for a 32-qubit 16-layer TFIM VQE task. The \texttt{DistributedContractor} achieves a $\mathbf{7.5\times}$ speedup on 8 GPUs compared to a single device, reducing the optimization step time from 17.86s to 2.38s.
    (b) Execution time per optimization step as a function of system size ($N \in [32, 40]$) on a fixed 8-GPU cluster. The scaling follows $T \propto 2^{1.1N}$, reflecting the combined complexity of Hilbert space dimension (N-qubit) and circuit depth ($L=N/2$). The framework successfully simulates a 40-qubit, 20-layer circuit with gradient for $11700$ circuit parameters in approximately 18 minutes per step.}
    \label{fig:h200_benchmarks}
\end{figure}

The results are summarized in Fig.~\ref{fig:h200_benchmarks}.
In terms of strong scalability (Fig.~\ref{fig:h200_benchmarks}a), the framework demonstrates exceptional efficiency. For a fixed 32-qubit workload, scaling from 1 to 8 GPUs reduces the gradient computation time from $17.86$s to $2.38$s, achieving a speedup of $7.5\times$. This indicates very small inter-GPU communication overhead for the implementation of \texttt{DistributedContractor}.
In terms of complexity scaling (Fig.~\ref{fig:h200_benchmarks}b), the system successfully simulates variational circuit training up to $N=40$ qubits and $L=20$ layers. The runtime scales as $T \propto 2^{1.1N}$, a manageable overhead compared to the theoretical scaling, confirming the effectiveness of the model parallelism engine in handling differentiable simulation for large quantum circuits.

\section{Integrated Demonstrations}

\subsection{Efficient ML Pipeline via Classical Shadows}

\begin{mdframed}
\textbf{Reference Script: }
\href{\rooturl examples/ng_whitepaper/VIIA_classical_shadow_ml.py}{\sf VIIA\_classical\_shadow\_ml.py}
\end{mdframed}

\textbf{Workflow}. A central challenge in QML is establishing an efficient data pipeline that connects quantum states to classical neural networks. Direct state tomography is exponentially expensive, while training on raw measurement counts often discards correlation information.
In this integrated example, we demonstrate a complete, end-to-end research workflow that bridges these domains using classical shadows~\cite{Huang2020shadow, Huang2021science}. The pipeline consists of three stages:
\begin{enumerate}
    \item \textbf{Physics Simulation:} We generate ground states of the TFIM across its quantum phase transition point ($g_c=1.0$).
    \item \textbf{Data Compression:} Instead of full tomography, we project the quantum states into classical shadows—compact classical descriptions composed of randomized Pauli measurements.
    \item \textbf{Machine Learning:} The raw shadow data (measurement bases and outcomes) are fed directly into a classical PyTorch neural network to learn a phase classifier (ordered vs. disordered) without explicit feature engineering.
\end{enumerate}

\textbf{Feature Highlight}.
To enable this pipeline at scale, \tc\ introduces a specialized \texttt{shadows} module designed for high-throughput data generation. Unlike standard frameworks that simulate measurements sequentially (shot-by-shot), \tc\ utilizes vectorized parallelism.
The \texttt{tc.shadows.shadow\_snapshots} function generates thousands of random Pauli bases and corresponding measurement bitstrings in a single, JIT-compiled tensor operation. This allows for the rapid construction of massive training datasets required for deep learning applications.

\begin{lstlisting}[language=Python]
# 1. Physics: Ground State Generation
# Solved via sparse linear algebra backend
psi = get_ground_state(n, g)

# 2. Data: Vectorized Shadow Generation
# Generate 20,000 snapshots in a single batch
# Pauli bases: 1=X, 2=Y, 3=Z
pauli_strings = np.random.randint(1, 4, size=(n_snapshots, n))

# The shadow engine handles the broadcasting and measurement simulation
snapshots = tc.shadows.shadow_snapshots(psi, pauli_strings, measurement_only=True)

# 3. Learning: PyTorch Integration
# The raw shadow data is tensor-compatible and fed into a generic NN
model, acc = train_model(X_shadows=snapshots, y_labels=phases)
\end{lstlisting}

\textbf{Benchmark}.
We benchmark the performance of this data generation pipeline by collecting $256$ snapshots for a 20-qubit system. The results (Table~\ref{tab:shadow_benchmark}) highlight the efficacy of \tc's tensor-native architecture.
By leveraging JAX's XLA compilation and GPU acceleration, \tc\ achieves a speedup of $\mathbf{10\times}$ compared to standard CPU-based simulation.

\begin{table}[h]
\centering
\caption{Time required to collect 256 classical shadow snapshots for a 20-qubit state. CPU is Apple M4 Pro and GPU is NVIDIA 5090 with \tc\ and Pennylane~\cite{Bergholm2018}.}
\label{tab:shadow_benchmark}
\begin{tabular}{lllc}
\toprule
\textbf{Framework} & \textbf{Backend} & \textbf{Hardware} & \textbf{Time (s)} \\
\midrule
PennyLane & Default & CPU & 6.50 \\
\tc & JAX & CPU & 2.85 \\
\textbf{\tc} & \textbf{JAX} & \textbf{GPU} & \textbf{0.29} \\
\bottomrule
\end{tabular}
\end{table}

\subsection{End-to-End Pure QML for Computer Vision}

\begin{mdframed}
\textbf{Reference Scripts: }
\href{\rooturl examples/ng_whitepaper/VIIB_cifar100_qml.py}{\sf VIIB\_cifar100\_qml.py}, 
\href{\rooturl examples/ng_whitepaper/VIIB_mnist_qml.py}{\sf VIIB\_mnist\_qml.py}
\end{mdframed}

\textbf{Workflow}.
While many QML demonstrations are limited to toy datasets like MNIST, \tc\ is engineered to handle realistic, high-dimensional machine learning tasks. This integrated example demonstrates an end-to-end pure quantum neural network pipeline for classifying images from the CIFAR-100 dataset. The workflow is adapted from code implementation of QML research~\cite{Chen2025plasticity, Chen2025resilience}.

The pipeline leverages several advanced features to manage the complexity of $32\times32$ color images (3072 features) and 100 output classes:
\begin{enumerate}
    \item \textbf{Amplitude Encoding:} We map the 3072-dimensional pixel vectors into the amplitudes of a 12-qubit state ($2^{12}=4096$). This provides an exponential compression of the input data.
    \item \textbf{Expressive Ansatz:} The circuit utilizes a brickwall generic $SU(4)$ two-qubit gates, maximizing the unitary expressive power per layer compared to standard CNOT-based ansatzes.
    \item \textbf{Readout Strategy:} To classify 100 classes, we perform a partial trace on the final state and use the probabilities of the first 7 qubits as the logit vector, mapping the first 100 outcomes to the target labels.
\end{enumerate}

\begin{lstlisting}[language=Python]
# 1. Amplitude Encoding (Data Loading)
# Input x is a normalized 4096-dim vector (padded 32x32x3 image)
c = tc.Circuit(N_QUBITS, inputs=x)

# 2. Deep Layer Construction with JAX Scan
# Instead of unrolling the loop, scan compiles it into a single efficient kernel
def one_layer(state, params):
    c_layer = tc.Circuit(N_QUBITS, inputs=state)
    # Apply SU(4) gates in a ladder pattern for max expressivity
    # Even pairs: (0,1), (2,3)...
    for i in range(0, N_QUBITS - 1, 2):
        c_layer.su4(i, i + 1, theta=params["even"][i // 2])
    # Odd pairs: (1,2), (3,4)...
    for i in range(1, N_QUBITS - 1, 2):
        c_layer.su4(i, i + 1, theta=params["odd"][(i - 1) // 2])
    return c_layer.state(), None

# Execute 10 layers efficiently
final_state, _ = jax.lax.scan(one_layer, c.state(), weights)

# 3. Readout (Partial Trace)
# Trace out last 5 qubits, keep first 7 (2^7 = 128 dims)
rho = tc.quantum.reduced_density_matrix(final_state, cut=range(7, 12))
probs = tc.backend.diagonal(rho)[:100] # Map to 100 classes
\end{lstlisting}

\begin{figure}[t]
    \centering
    \includegraphics[width=0.96\linewidth]{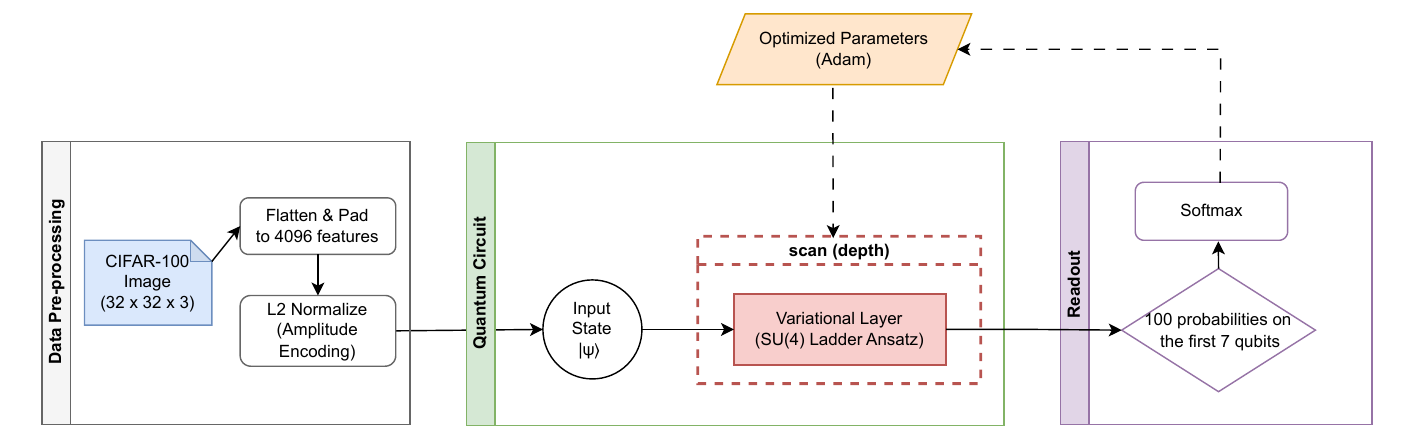}
    \caption{\textbf{Pipeline of the End-to-End QML Pipeline for CIFAR-100.} 
    The pipeline utilizes amplitude encoding for high-dimensional input ($32\times32$ pixels, 3 channels), processes data via a JAX-compiled deep quantum circuit with expressive SU(4) gates, and performs classification via probability on subset qubits. The entire workflow is differentiable and optimized end-to-end.}
    \label{fig:cifar100_arch}
\end{figure}

\textbf{Feature Highlight}.
A key innovation in this example is the use of \texttt{jax.lax.scan}. In standard frameworks, a circuit with $L$ layers is unrolled during compilation, creating a computational graph that grows linearly with depth. For very deep quantum neural networks, this leads to excessive compilation times.
By wrapping the layer logic in \texttt{scan}, \tc\ compiles the circuit into a fixed-size loop kernel. This reduces the compilation time from minutes to seconds and allows the memory footprint to remain constant regardless of depth, enabling the training of extremely deep quantum models on GPU hardware. If the scan kernel is further decorated with \texttt{jax.checkpoint}, the memory cost can also be further reduced.

\textbf{Benchmark}.
To validate the scalability of the QML pipeline, we benchmarked the training throughput on both CPU and GPU hardware. The results, summarized in Table~\ref{tab:qml_benchmark}, demonstrate the efficiency of the JAX-based tensor backend of TensorCircuit-NG.
For the 10-class MNIST task (10 qubits, 40 layers), the RTX 5090 achieves a training speed of 1.88 seconds per epoch for the full 60,000-sample dataset, delivering an approximate $\mathbf{8\times}$ speedup over the CPU baseline. In the more computationally demanding CIFAR-100 task (12 qubits, 30 layers), the GPU advantage becomes even more pronounced. Processing large batches of 5,000 high-dimensional inputs, the RTX 5090 completes an epoch with a speedup of over $\mathbf{11\times}$ compared to the CPU. This capability is pivotal for probing the boundaries of potential quantum advantage in QML.

\begin{table}[h]
\centering
\caption{QML training performance benchmark (training time per epoch). The pipeline leverages JIT compilation and vectorized batching provided by JAX with single precision. CPU refers to Apple M4 Pro and GPU refers to NVIDIA RTX 5090.}
\label{tab:qml_benchmark}
\begin{tabular}{lccccccc}
\toprule
\textbf{Task} & \textbf{Qubits~} & \textbf{Layers~} & \textbf{Data Size~} & \textbf{Batch Size~} & \textbf{\# of Params} &\textbf{CPU} & \textbf{GPU} \\
\midrule
MNIST & 10 & 40 & 60,000 & 15,000 & 5400& 15.00 s~ & \textbf{1.88 s} \\
CIFAR-100 & 12 & 30 & 50,000 & 5,000 & 4950&53.15 s~ &  \textbf{4.60 s} \\
\bottomrule
\end{tabular}
\end{table}

\subsection{Universal Quantum Dynamics of Haar-Random MIPT}
\begin{mdframed}
\textbf{Reference Script: }
\href{\rooturl examples/ng_whitepaper/VIIC_mipt_prob.py}{\sf VIIC\_mipt\_prob.py}
\end{mdframed}

\textbf{Workflow}.
While stabilizer formalisms (Section V.B) allow for simulating thousands of qubits, they are restricted to the discrete Clifford group and cannot capture universal quantum chaotic behavior or fine-grained entanglement spectra. To study generic phase transitions—such as the MIPT in Haar-random circuits, one must simulate the full wavefunction evolution.
This example demonstrates \tc's capacity for simulating universal quantum dynamics, leveraging JAX's control flow primitives to efficiently simulate thousands of stochastic trajectories in parallel.

The simulation targets a one-dimensional chain of $N$ qubits. The system evolves under a brick-wall circuit architecture composed of $D$ discrete time steps. Each time step consists of:
(1) \textbf{Unitary Evolution:} Two layers of nearest-neighbor two-qubit gates acting on even and odd bonds respectively. These gates are drawn independently from the Haar measure on $SU(4)$, ensuring maximal scrambling of quantum information.
(2) \textbf{Probabilistic Measurement:} A layer of single-qubit projective measurements in the computational basis ($Z$-basis). Each qubit is measured with probability $p$ (collapsing the wavefunction) and remains unperturbed with probability $1-p$.

For the performance benchmarks presented below, we configure the system with $N=20$ qubits and $D=40$ layers to evaluate the simulator's throughput in the regime of deep, entangled dynamics.

\textbf{Feature Highlight}.
Simulating MIPT requires evolving an ensemble of wavefunctions under random unitary gates and stochastic projective measurements. This presents two software challenges: dynamic control flow (measurements are random) within a static computational graph and massive sampling requirements. \tc\ addresses these via a robust JAX integration:

\begin{itemize}
    \item \textbf{Deep Circuit Optimization with \texttt{scan}:}
    Standard circuit unrolling creates a massive computational graph for deep evolution, leading to slow compilation. We utilize \texttt{jax.lax.scan} to treat the time evolution as a loop. This reduces the compilation complexity from $O(D)$ to $O(1)$, keeping the executable binary small and cache-friendly even for long-time dynamics.

    \item \textbf{Parallel Trajectory Batching with \texttt{vmap}:}
    Instead of serial execution, we use \texttt{jax.vmap} to vectorize the simulation over the batch dimension. This allows a single GPU to process thousands of unique quantum trajectories simultaneously, saturating the massive arithmetic throughput of modern accelerators.

    \item \textbf{Jittable Measurement:}
    The measurement process is implemented via \texttt{general\_kraus}, which supports tracking the probability of each trajectory while being compatible with static computational graph for JIT and AD.
\end{itemize}

\textbf{Benchmark}.
We benchmarked the throughput of this universal simulation pipeline on a 20-qubit, 40-layer Haar-random circuit using single precision (\texttt{complex64}). The task involves evolving a batch of random trajectories and computing their final probability for the trajectory according to Born's rule.
As shown in Table~\ref{tab:mipt_benchmark}, the combination of \texttt{scan} and \texttt{vmap} on the NVIDIA H200 GPU achieves an effective throughput of just \textbf{84 milliseconds per trajectory} for this computationally intensive 20-qubit 40-layer task. This efficiency enables the accumulation of high-resolution statistics for phase transition analysis in a practical timeframe.

\begin{table}[h]
\centering
\caption{MIPT Simulation Performance (20 Qubits, 40 Layers). Effective time denotes the amortized wall-clock time to simulate a single trajectory within a large batch. CPU refers to Apple M4 Pro and GPU refers to NVIDIA H200.}
\label{tab:mipt_benchmark}
\begin{tabular}{lccc}
\toprule
\textbf{Hardware} & \textbf{Batch Size~} & \textbf{Total Time (s)~} & \textbf{Effective Time / Traj} \\
\midrule
CPU & 1000 & 1097 s & 1.097 s \\
GPU & 1000 & \textbf{84.16 s} & \textbf{0.084 s} \\
\bottomrule
\end{tabular}
\end{table}

\subsection{Variational Simultaneous Excited States with MPS}

\begin{mdframed}
\textbf{Reference Module: }
\href{https://github.com/sxzgroup/quantum_excited_state/tree/main/scripts/mps_example}{\sf mps\_example}
\end{mdframed}

\textbf{Workflow}.
Solving for the low-energy spectrum of many-body systems is a grand challenge in computational physics. \tc\ enables a novel variational framework that simultaneously targets multiple excited states without enforcing explicit orthogonality~\cite{Zhang2025es}. The method minimizes the trace  $L = \text{Tr}(S^{-1}H)$, where $S_{ij} = \langle \psi_i | \psi_j \rangle$ and $H_{ij} = \langle \psi_i | \hat{H} | \psi_j \rangle$ are the overlap and Hamiltonian matrices constructed from a set of $N_s$ non-orthogonal, parameterized states.

\begin{figure}[t]
    \centering
    \includegraphics[width=0.7\linewidth]{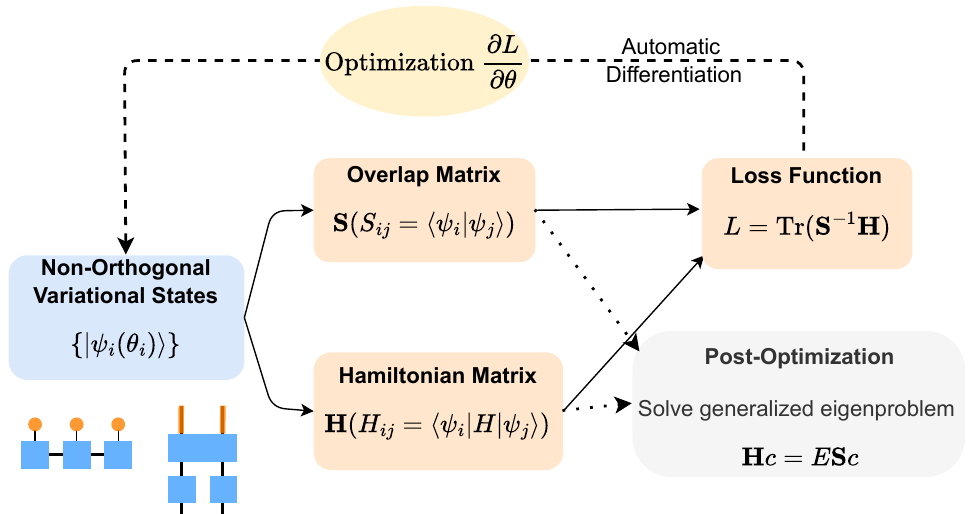}
    \caption{\textbf{End-to-End Variational Workflow for Excited States.} 
    The framework operates by constructing the overlap ($S$) and Hamiltonian ($H$) matrices from a set of parameterized, non-orthogonal states $\{|\psi_i(\theta)\rangle\}$ which can be built on top of tensor networks, neural networks or quantum circuits. 
    The variational parameters are optimized by minimizing the universal loss function $L = \text{Tr}(S^{-1}H)$ using automatic differentiation gradients $\frac{\partial L}{\partial \theta}$. 
    Finally, the approximate low-energy spectrum is retrieved by solving the generalized eigenvalue problem $Hc=ESc$ in the post-optimization phase. Reproduced from Ref.~\cite{Zhang2025es}.}
    \label{fig:qes_workflow}
\end{figure}

\textbf{Feature Highlight.} 
This integrated example showcases a high-performance implementation for 1D spin chains using periodic MPS. The pipeline epitomizes the unified platform philosophy:
\begin{enumerate}
    \item \textbf{Initialization (Quimb):} We leverage the ecosystem by using \texttt{quimb} to construct the periodic Heisenberg Hamiltonian MPO, which is then seamlessly converted into a \tc\ tensor format.
    \item \textbf{Contraction (TensorCircuit):} We define custom contraction routines for the overlap ($S_{ij} = \langle \psi_i | \psi_j \rangle$) and energy matrices ($H_{ij} = \langle \psi_i | \hat{H} | \psi_j \rangle$). Using \texttt{vmap}, we batch these contractions over the entire subspace of $k$ states.
    \item \textbf{Optimization (Optax):} The complex loss function is fed into GPU accelerated L-BFGS optimizer from the JAX ecosystem, with the entire step JIT-compiled.
\end{enumerate}

This application explicitly demonstrates that \tc\ is not solely a quantum circuit simulator. It functions as a high-performance, general-purpose tensor network engine capable of supporting complex quantum many-body calculations.
The script utilizes custom contraction patterns  that go beyond standard gate operations, proving the platform's flexibility in handling arbitrary tensor network structures.
Moreover, the entire calculation is end-to-end differentiable.

\begin{lstlisting}[language=Python]
# 1. Unified Ecosystem: Import Hamiltonian from Quimb
# Construct Heisenberg MPO using Quimb's robust physics modules
qb_mpo = quimb.tensor.tensor_builder.MPO_ham_heis(L, j=1.0, cyclic=True)
# Seamlessly convert Quimb tensors to TensorCircuit backend (JAX tensors)
h_mpo = tc.backend.stack([qb_mpo.tensors[i].data for i in range(L)])

# 2. Differentiable Tensor Network Engine
# Use vmap to vectorize the contraction over the N_s * N_s subspace
# Transforming single-pair contraction logic into batch matrix construction
compute_S_vmap = K.jit(K.vmap(contract_overlap, vectorized_argnums=(0, 1)), static_argnums=2)
compute_H_vmap = K.jit(K.vmap(contract_hamiltonian, vectorized_argnums=(0, 1)), static_argnums=3)

@K.jit
def loss(params, h_mpo, N):
    # Compute dense S and H matrices for the non-orthogonal subspace
    bras, kets = prepare_batch_pairs(params)
    S = K.reshape(compute_S_vmap(bras, kets, N), (k, k))
    H = K.reshape(compute_H_vmap(bras, kets, h_mpo, N), (k, k))
    
    # Universal Loss: Minimize sum of Generalized Eigenvalues
    # Gradients flow through matrix inversion S^{-1} and tensor contraction
    return K.real(K.trace(K.inv(S + eps * K.eye(k)) @ H))

# 3. High-Performance Optimization (JIT + L-BFGS)
solver = optax.lbfgs() # Use second-order optimizer for faster convergence

@K.jit
def optimization_step(params, opt_state):
    value, grads = K.value_and_grad(loss)(params, h_mpo, L)
    
    # L-BFGS update step (fully compiled via XLA)
    updates, opt_state = solver.update(
        grads, opt_state, params, value=value,
        value_fn=partial(loss, hs=h_mpo, L=L) # Line search requires value function
    )
    return optax.apply_updates(params, updates), opt_state, value
\end{lstlisting}

\textbf{Benchmark.}
We benchmark the performance of this variational solver on an NVIDIA H200 GPU. The task involves optimizing a subspace of $N_s=16$ excited states of different system sizes for 1D Heisenberg chains of varying lengths $L$, using a bond dimension of $\chi=16$.

As shown in Table~\ref{tab:mps_benchmark}, \tc\ demonstrates exceptional scalability. The time per optimization step scales near linearly with system size $L$, enabling the study of large-scale quantum systems ($L=128$) with rapid iteration times, making it a powerful tool for discovering many-body physics.

\begin{table}[h]
\centering
\caption{Average execution time per L-BFGS optimization step for finding 16 excited states (bond dimension $\chi=16$) on an NVIDIA H200 GPU with double precision.}
\label{tab:mps_benchmark}
\begin{tabular}{ccc}
\toprule
\textbf{System Size~} &  \textbf{~Time per Step (s)} \\
\midrule
32 & 0.31 \\
64 &  0.74 \\
128 &  1.82 \\
\bottomrule
\end{tabular}
\end{table}

\subsection{Broad Application Gallery}

The examples detailed above represent only a fraction of the capabilities hosted in the \tc\ ecosystem. The ecosystem includes a rich collection of modules and scripts covering diverse advanced topics, ensuring researchers can find starting points for almost any task. Notable implementations include:

\begin{itemize}
    \item \textbf{Quantum-Classical Hybrid Reinforcement Learning:} A seamless integration of JAX-based quantum circuits simulation with the \texttt{Stable Baselines3} framework. This allows for the training of hybrid agents with end-to-end differentiability across PyTorch and Jax frameworks~\cite{Chen2025plasticity}.~\fancylink{https://github.com/sxzgroup/quantum-plasticity/blob/main/src/rl.py}{Reference Script: \texttt{rl.py}}.
    \item \textbf{Pauli Propagation and Operator Dynamics:} Efficient approximate simulation in Heisenberg picture of operator spreading and quantum circuits~\cite{Shao2024, Angrisani2025} using JIT-compiled and differentiable Pauli propagation, enabling the study in the context of VQAs~\cite{Li2025lwpp}.~\fancylink{https://github.com/ZongliangLi/lwpp_init}{Reference Module: \texttt{lwpp}}. 
    \item \textbf{Variational Imaginary Time Evolution:} Robust solvers for simulating imaginary time dynamics on parameterized quantum circuits, essential for ground state preparation and thermal state modeling in many-body physics~\cite{Yuan2019, Zhang2023vite}.~\fancylink{https://github.com/tensorcircuit/tensorcircuit-ng/blob/master/examples/variational_dynamics.py}{Reference Script: \texttt{variational\_dynamics.py}}.
    \item \textbf{Unified Variational Optimization and Evaluation:} A unified framework to optimize and compare diverse ansatz architectures. Researchers can benchmark neural quantum states, parameterized quantum circuits, MPS, and projected entangled pair states on the same Hamiltonian, utilizing consistent metrics such as the effective temperature description to evaluate the quality of variational approximations~\cite{Chen2025et}.~\fancylink{https://github.com/sxzgroup/et}{Reference Module: \texttt{et}}.
\end{itemize}

This extensive library of examples serves not just as documentation, but as a validated codebase that accelerates the transition from theoretical concept to numerical realization.

\section{Conclusion and Outlook}

\tc\ marks a definitive transition in the quantum software landscape, evolving from a standalone simulation tool into a comprehensive computational infrastructure for the AI and HPC era. By strictly adhering to a tensor-native philosophy, we have successfully unified diverse physical simulation engines under a single, differentiable programming interface. This architecture not only democratizes access to high-performance GPU/TPU acceleration but also enables the seamless composition of quantum modules with state-of-the-art classical machine learning pipelines.

The benchmarks presented, particularly the simulation of 40-qubit variational quantum circuit systems and high-dimensional QML tasks, demonstrate that Python-based frameworks can achieve superior performance when architected correctly on top of XLA and distributed runtimes.

Looking forward, the development of \tc\ will target the following strategic frontiers: \begin{enumerate} \item \textbf{AI for Quantum Physics:} Leveraging the differentiable physics engine to power the next generation of generative models for quantum state reconstruction, Hamiltonian learning, ansatz discovery and pulse control.

\item \textbf{Towards Fault-Tolerance:} Utilizing the unified simulation paradigm—combining large-scale stabilizer simulations with tensor network contraction—to design and validate novel quantum error correction codes and toolkits. 

\item \textbf{Expanded Hardware Integration:} Building upon our preliminary infrastructure for QPU access, we aim to further bridge the gap between numerical simulation and experimental execution. Future work will focus on advanced compilation techniques, hardware-aware error mitigation strategies, and broadening connectivity to a diverse range of quantum hardware providers. \end{enumerate}

 \tc\ is positioned not merely as an academic research tool, but as a robust, industrial-grade infrastructure committed to open-source principles. Backed by a dedicated team of core developers and a growing community from leading research institutions, we are committed to providing long-term support, ensuring API stability, and continuously integrating the latest advancements in AI and quantum hardware. This reliability makes \tc\ a secure and strategic technology choice, serving as a dependable foundation for building the next generation of quantum software stacks and driving the next wave of discovery in quantum science.

 \section*{Acknowledgments}
 This work is supported by Quantum Science
and Technology-National Science and Technology Major Project (No. 2024ZD0301700), the National Natural Science Foundation of China (Nos. 12504599 and 12574546), and NSAF (No. U2330401).
 
%

\end{document}